\documentclass{article}

\usepackage{arxiv}

\usepackage[utf8]{inputenc} % allow utf-8 input
\usepackage[T1]{fontenc}    % use 8-bit T1 fonts
\usepackage{hyperref}       % hyperlinks
\usepackage{url}            % simple URL typesetting
\usepackage{booktabs}       % professional-quality tables
\usepackage{amsfonts}       % blackboard math symbols
\usepackage{nicefrac}       % compact symbols for 1/2, etc.
\usepackage{microtype}      % microtypography
\usepackage{lipsum}
\usepackage{graphicx}
\graphicspath{ {./images/} }

\usepackage{amssymb}
\usepackage{multirow}
\usepackage{xcolor} 
\usepackage{amsmath}
\usepackage[normalem]{ulem}
\usepackage{caption}
\usepackage{subcaption}
\usepackage{colortbl}
\usepackage{rotating}
\usepackage{soul}

\title{Optimizing 2D Input Representations and Sub-phase Fusion Strategies for Differential Diagnosis of Asthma and COPD Using CNN- and GRU-Based Networks}

\author{
İpek Şen \\
  Dept. Electrical and Electronics Engineering\\
  Istanbul Bilgi University, Turkey\\
  \texttt{ipek.sen@bilgi.edu.tr} \\
  \And
  Özgür Özdemir\\
  Dept. Computer Engineering\\
  Istanbul Bilgi University, Turkey\\
  \texttt{ozgur.ozdemir@bilgi.edu.tr} \\
  \And
  Elena Battini Sönmez\\
  Dept. Computer Engineering\\
  Istanbul Bilgi University, Turkey\\
  \texttt{elena.sonmez@bilgi.edu.tr} \\
}

\begin{document}
\maketitle
\begin{abstract}
The vector auto-regressive (VAR) model demonstrated success in conveying useful spatio-temporal information in multi-channel pulmonary sound measurements for the differential diagnosis of asth-ma and chronic obstructive pulmonary disease (COPD). This study aims to explore the performance of the VAR model in comparison with spectrogram-based signal representations using deep learning. In pulmonary sound classification, spectrogram-based representations suffer from inconsistent temporal dimensions due to varying respiratory cycle durations. Testing two alternative methods to equate their temporal dimensions, optimizing their spectral and temporal dimensions for the diagnosis, and testing the methods to fuse the respiratory sub-phases (early, mid, and late inspiration and expiration) for the best performance are also included in the objectives. In addition to the VAR matrices, mel-frequency cepstral coefficient (MFCC) matrices and log-mel spectrograms are computed from 14-channel pulmonary sounds of 50 subjects (30 asthma, 20 COPD). Along with traditional trimming/zero-padding, adaptive-length windowing was presented to fix their temporal dimensions. Their spectral and temporal dimensions were optimized by testing a range of parameters. Different convolutional neural network (CNN) architectures were employed to extract features from the two-dimensional representations obtained over the sub-phases. The extracted sub-phase features were then fused using various strategies including direct concatenation, gated recurrent unit (GRU) network and GRU with attention mechanism. Model performances were assessed through respiratory cycle-based evaluation and subject-based evaluation comprising multiple respiratory cycles. Several data augmentation techniques were also studied to cope with limitations in data size. The best cycle-based F1-score (0.877) was obtained using the MFCC matrices with thirteen coefficients and 64-point time resolution per sub-phase representation followed by direct feature concatenation, and the best subject-based F1-score (0.855) was obtained using the MFCC matrices with thirteen coefficients and 256-point time resolution per full-cycle representation, both obtained by adaptive-length windowing. Augmentation degraded the performance of models overall, yet mixup augmentation was the best among the methods tested. MFCC outperformed log-mel spectrogram and VAR model in differentiation of asthma and COPD. Sophisticated fusion strategies did not improve the diagnosis. Augmentation did not contribute, demonstrating the significance of authentic data in pulmonary sound studies.
\end{abstract}

% keywords can be removed
%\keywords{Asthma \and COPD \and VAR \and MFCC \and log-mel spectrogram \and CNN \and GRU \and adaptive-length windowing \and trimming-padding }

\section{Introduction}
\label{sec:Intro}

In clinical practice, a clear distinction between asthma and chronic obstructive pulmonary disease (COPD) may be difficult due to the overlapping symptoms, therefore misdiagnosis is not rare \cite{Tinkelman2006}, \cite{Bellia2003}, \cite{VanSchayck1996}, \cite{VanWheel2002}. Although some discriminative symptoms and characteristics are described in the medical literature (in terms of, e.g., reversibility of airway obstruction, persistency of coughing, existence of wheezing and allergy, sputum production, age, smoking, and speed of progression \cite{Bellia2003}, \cite{VanSchayck1996}, \cite{Burrows1991}, \cite{ATS1987}), it is still difficult to make a clear distinction since: (i) an overlap between these characteristics and symptoms still exists \cite{Tinkelman2006}, \cite{Bellia2003}, and (ii) the diagnostic power of many of them depends on how accurately the patient reports his/her lifestyle and complaints. The only objective distinction can be done through a pulmonary function test called spirometry, however, the success of the assessment depends strongly on how well the patient performs the test, which may be difficult depending on his/her age or physical condition \cite{hangaard2017causes}, \cite{pistelli2011practical}. Moreover, spirometry testing and assessment are not available at all levels of healthcare.

Pulmonary sounds contain valuable information, however stethoscope auscultation is subjective by nature since it depends on the hearing ability and experience of the physician, and it does not allow quantification and objective comparison in conventional medical practice. On the other hand, computerized sound analysis techniques render it possible to quantify the acoustic information towards differential diagnosis of pulmonary conditions. The earliest study in the literature comparing asthma and COPD through quantitative analysis of pulmonary sounds reports that the spectral differences between asthma and COPD can hardly be detected via stethoscope auscultation, however, they can be objectively quantified via computerized sound analysis techniques \cite{Malmberg1995}.

Although the study in \cite{Malmberg1995} showed that differences exist between asthma and COPD sounds, it did not suggest a computerised method that can be used in clinical practice for the differential diagnosis (e.g., classification). Five later studies \cite{khodabakhshi2017}, \cite{Islam2018}, \cite{haider2022}, \cite{koshta2023}, \cite{Sen2021} proposed classification algorithms for differentiating asthma and COPD, the first four including also the healthy group as the third class, whereas the last one focusing only on the differential diagnosis of the two diseases. No additional studies specifically focusing on differentiating asthma and COPD using pulmonary sounds have been identified in the existing literature. To better situate the present study within the literature, the following paragraphs present information on these five studies with particular attention to details such as the datasets used, the number of subjects and channels, the duration and characteristics of the signal segments from which the features were calculated, and whether channel and/or segment fusion was performed.

The study in \cite{khodabakhshi2017} used six-channel pulmonary sounds of 83 subjects (27 COPD, 31 asthma, 25 healthy) recorded at the Shariati Hospital, Tehran. Multiple types of features were employed: AR coefficients, wavelet transform features, mel-frequency cepstral coefficients (MFCC), temporal modelling features via attractor recurrent neural network (temp-ARNN), those via ARNN with fuzzy functions (temp-FFsARNN), recurrence quantification analysis (RQA) features, RQA features via ARNN (RQA-ARNN), and those via ARNN with fuzzy functions (RQA-FFsARNN). For classification, a three-layer neural network was used in four-fold cross validation scheme. A separate model was trained for each channel and the overall decision was made via majority voting over the six channel decisions. RQA-FFsARNN was found to attain the best performance with an overall three-class classification accuracy of 91\%. The study used only one \textit{full-cycle} (the period from the beginning of an inspiration to the end of the subsequent expiration, will be called one \textit{full-cycle} hereafter) from each subject. Inspiration-expiration periods were recognized by listening to the sounds. Although the study used the entire full-cycle for the calculation of time-series sequences, fixed-length windowing was performed using 500 ms-long windows with 20 ms overlap for the calculation of RQA features, then the features were averaged over the windows. No details were given for how MFCC were calculated, including the window length, the overlap between the successive windows, window type, number of FFT points, number of triangular mel filters, or the coefficients kept after the DCT.

The study in \cite{Islam2018} used four-channel pulmonary sounds recorded at the Institute of Pulmocare and Research, Kolkata, from 60 subjects (20 in each of the COPD, asthma and healthy groups) during one full-cycle. Power spectral density (PSD) was estimated and statistical features (mean of the absolute value, variance, kurtosis and skewness) were calculated from spectral sub-bands for each channel. The feature vectors obtained from the four channels were fed into the classifiers both individually and after being concatenated to form a single feature vector associated with the subject. A three-layer neural network was used in leave-one-(subject-)out cross validation scheme, and the overall cross-validation was repeated 25 times to account for variability due to random initialization. The highest performance was obtained when the channels were combined, yielding a three-class classification accuracy of 60.33\%. In the study, the use of PSD resolved the issue of varying full-cycle durations; however, it did not allow for the utilization of time-varying spectral information. 

The study in \cite{haider2022} used single-channel pulmonary sounds recorded from 240 subjects (80 in each of the COPD, asthma and healthy groups) at All India Institute of Medical Sciences, Raipur. Wavelet entropy and wavelet packet energy were calculated from the entire 20-second recordings and were fed into four different classifiers, namely, SVM, decision tree (DT), k-nearest neighbor (kNN), and discriminant analysis (DA). The DT classifier with 33\% holdout validation was reported to achieve the best performance with an overall (three-class) accuracy of 99.3\%. The corresponding class accuracies were also reported: 97.6\% and 100\% for asthma and COPD, respectively.

The study in \cite{koshta2023} used the single-channel pulmonary sound data of 76 subjects (9 COPD, 32 asthma, 35 healthy) from a publicly available dataset \cite{fraiwan2021dataset} recorded at King Abdullah University Hospital, Ramtha. Three recordings (in bell, diaphragm, and extended filtration modes) were taken from each subject, making 228 sound recordings in total. Every subject was auscultated at one of ten predefined chest locations, with sites varying across individuals. The recordings were divided into 5-second segments (without paying particular attention to their timing in relation to breathing cycles), then Fourier Bessel Series Expansion (FBSE) features were calculated from those segments, and were fed into various machine learning classifiers and a custom-designed one-dimensional convolutional neural network (1D-CNN). The highest accuracies were obtained using ensemble subspace kNN (94.1\%) at level one (normal vs. abnormal classification) via 10-fold cross validation scheme, and using ensemble bagged tree (96.6\%) at level two (asthma vs. COPD classification) via both 10-fold and 5-fold cross validation schemes. The 1D-CNN network yielded accuracies of 96.4\% and 81.1\% at levels one and two, respectively. The reported accuracies were segment-based, more precisely, any segment fusion at the level of features or decisions to derive subject-based diagnoses was not reported. Furthermore, the possibility that data from the same subject or recording appeared both in the training and validation sets was not explicitly ruled out. Ultimately, it remains unconvincing that the classifier achieves fair learning, considering the variability across filtration modes and recording sites.

The study in \cite{Sen2021} used several full-cycles of 14-channel pulmonary sounds by further dividing each full-cycle into its \textit{sub-phases} (each one of the early, mid, and late phases of inspiration and expiration will be called a \textit{sub-phase} hereafter). Channel fusion was automatically performed by the multivariate model used to calculate the features, namely, the vector auto-regressive (VAR) model. First, every sub-phase was further divided into 250-point segments to meet the stationarity assumption of the model, and a matrix of model coefficients was estimated for each such segment. Then, the matrices were z-scanned into vectors (were vectorized) and fed into the Gaussian mixture model (GMM) classifier in leave-one-subject-out cross validation scheme. Segment decisions were combined to obtain the sub-phase decisions, and sub-phase decisions were  combined to obtain the subject-decisions (diagnostic predictions). The study used pulmonary sounds recorded from 50 subjects (20 COPD and 30 asthma) at Yedikule Hospital, Istanbul (see Section \ref{sec:data} for details). A binary classification accuracy of 98\% was achieved (only one COPD subject being misdiagnosed).

The work presented here is a continuation of the study in \cite{Sen2021} and aims to: 
\begin{itemize}
    \item investigate the ability of the VAR model (to differentiate asthma and COPD) using 2D-CNNs, and
    \item compare the performance of the VAR model (in differentiating asthma and COPD) with that of the log-mel spectrogram and MFCC,
\end{itemize}
which have not been performed in the pulmonary sounds analysis literature so far. Additional novel contributions introduced in the course of fulfilling these two aims are outlined as follows:

\begin{itemize}
\item The two-dimensional input representations must be of the same size to be fed into CNNs. However, the sizes of log-mel spectrogram and MFCC matrix depend on signal and window lengths, unlike the VAR model coefficient matrices. Accordingly, the present work: 
\begin{itemize}
\item formulates and compares two different approaches to fix the temporal sizes of the spectro-temporal input representations computed for unequal durations of inspiration, expiration, and/or sub-phases, and 
\item optimizes the temporal and spectral resolutions of those representations by performing grid search over a set of parameters.
\end{itemize}
\item The present work divides the sounds into sub-phases as in \cite{Sen2021} to utilize the timing information of acoustic events over the breathing cycle. However, since CNNs are now being used instead of classical machine learning classifiers, it proposes and compares novel sub-phase fusing strategies involving dedicated network design and input preparation.
\end{itemize}

CNNs were originally designed to process images; therefore, it is common to use two-dimensional representations (e.g., spectrograms) as inputs to CNNs, although the data being classified is one-dimensional (e.g., sound signal). Additionally, while RGB images provide a three-channel input (red, green, and blue) to a CNN classifier, CNN can be extended to accept a higher number of channels, hence making it suitable for multi-channel pulmonary sound measurements. Although newer methods are emerging, it is still a common approach in pulmonary sound classification to use CNN architectures with two-dimensional (spatially structured) input representations such as short-time Fourier transform (STFT), MFCC, chromagram, spectrogram, or mel spectrogram; either to classify pulmonary conditions (i.e., \textit{diagnostic classification}, as referred to hereafter) \cite{Srivastava2021}, \cite{zhang2024}, \cite{shehab2024} as in this study, or to classify particular waveforms (i.e., adventitious sounds) \cite{jung2021, kim2021, asatani2021, zulfiqar2021, petmezas2022, wang2024, kim2025}. The two diagnostic classification studies addressed here classified COPD vs. non-COPD (multiple diseases) \cite{Srivastava2021}, or various chronic and non-chronic diseases along with the healthy class in three-class and six-class classification schemes \cite{zhang2024}, using custom-designed CNNs; whereas the third study \cite{shehab2024} classified eight conditions including seven diseases along with the healthy class using pretrained CNN models (DenseNet201, InceptionResNetV2, MobileNetV2, ResNet101, and EfficientNetB0). In this study, a variety of large-scale CNN architectures, namely ResNet \cite{he2016}, Wide-ResNet \cite{zagoruyko2016}, DenseNet \cite{huang2017}, and VGG \cite{simonyan2014}, as well as a custom-designed CNN network were employed for feature extraction (from full-cycles or sub-phases); and Gated Recurrent Units (GRU) \cite{bahdanau2014} were used for feature (sub-phase) fusion. Temporal convolutional network (TCN) \cite{bai2018} was also tested for feature extraction.

MFCC has been successfully used for audio and speech processing, and also proved to be effective in pulmonary sound analysis \cite{khodabakhshi2017, Srivastava2021, zhang2024, jung2021, wang2024, kim2025, sengupta2016, palaniappan2014, haider2019}. In \cite{Srivastava2021}, various representation types were compared and the MFCC matrix achieved the best classification performance, outperforming the chromagram and mel spectrogram. Traditionally, 13 coefficients have been used in speech processing, therefore most of the pulmonary sound studies also used 13 coefficients \cite{jung2021, kim2025, palaniappan2014, haider2019}. However, other values such as 10 \cite{jung2021}, 15 \cite{wang2024}, 20 \cite{zhang2024, jung2021, sengupta2016} and 40 \cite{Srivastava2021} were also adopted. The number of mel filters and MFCC was the same in all these studies except for \cite{wang2024} and \cite{kim2025}, in which 20 \cite{wang2024} and 40 \cite{kim2025} filters were applied and 15 \cite{wang2024} and 13 \cite{kim2025} coefficients were retained for further use. Typically, the studies used a fixed set of calculation parameters (e.g., coefficient count, window length, or overlap ratio) chosen heuristically, and did not optimize them for the particular application. The studies in \cite{jung2021, wang2024, sengupta2016} performed such optimization; however, they did so specifically for adventitious sound classification. In most diagnostic classification studies, aside from absence of optimization, little \cite{Srivastava2021, zhang2024, palaniappan2014, haider2019} to no information \cite{khodabakhshi2017} was given about the parameters used in the calculations. In fact, for comparison and/or full understanding of methodologies, the following parameters should be reported: (i) the length of the signal segments for which the spectrograms (the first step of MFCC calculation) are computed, or some description that can replace the length, e.g., entire inspiration, (ii) length of the sliding windows applied, (iii) their overlap ratio, (iv) number of FFT points, (v) number of triangular mel filters applied on the spectrogram to obtain the mel spectrogram, and (vi) number of coefficients retained after the discrete cosine transform (DCT) (the last step of MFCC calculation), i.e., MFCC count. These parameters effect the temporal and spectral resolutions, hence the discriminative power of the 2D representations. 

To equalize the temporal sizes of the spectro-temporal representations before the classifier, two main approaches have been adopted in the literature. The study in \cite{Srivastava2021} (using MFCC and mel spectrogram) trimmed the longer signals, and zero-padded the shorter ones, to 20 seconds (the duration of the recordings ranged from 10 to 90 seconds) to equalize the signal length before calculating the spectrogram over the entire signal. The study in \cite{Messner2020} (using log-spectrogram) used one full-cycle per subject and zero-padded each to match the duration of the longest cycle within the set of all subjects. On the other hand, in \cite{zhang2024} ``a range of 20-40 ms" is reported for the sizes of the sliding windows, which implies that different (\textit{adaptive}, as will be mentioned in this study) window lengths might have been applied to achieve equal temporal sizes of matrices from recordings of different durations. 

In this study, mainly the MFCC matrices and log-mel spectrograms were used for comparison with the VAR model matrices. These representations were computed from sub-phases as in \cite{Sen2021} and also from full-cycles as in \cite{khodabakhshi2017, Islam2018, Messner2020}; however, multiple such cycles were used as was done only in \cite{Sen2021}. Two different approaches (trimming/padding and adaptive-length windowing) for equalizing the temporal sizes of these representations were compared, where trimming and padding were done according to the shortest and longest full-cycle as in \cite{Messner2020}. Moreover, various window sizes were experimented with an aim to optimize the resulting temporal resolution. In MFCC calculation, all coefficients were retained after the DCT, i.e., the number of MFCC was equal to the number of triangular mel filters applied, implying equal spectral sizes for the two types of representations. Mel filter, hence MFCC, counts of 13, 26, and 39 were experimented with, to include the most frequently used value of 13 and its two- and three-fold multiples, thereby covering the range of typical values adopted in the pulmonary sound analysis literature. The log-spectrogram was only tested in one experiment because it was not possible to use it in all experiments due to its much higher spectral resolution compared to the log-mel spectrogram.

Timing of the acoustic changes caused by the disease (whether they are heard in the early, mid or late inspiration or expiration) plays an important role in the assessment of acoustic information in stethoscope auscultation \cite{Gavriely1995}, \cite{Nath1974}, \cite{Sovijarvi2000p591}, \cite{Douros2018}, \cite{Piirila1991}, hence in computerized analysis of pulmonary sounds. Segmenting the signal in relation to full-cycles and sub-phases makes it possible to retain and exploit this timing information for improved diagnostic success, as demonstrated in \cite{Sen2021}. Part of the experiments in this study were conducted to compare the proposed strategies for fusing the sub-phases to derive the cycle-based decisions, over which the subject-based decisions (diagnostic labels) were predicted by simple majority voting. Producing a single diagnostic label per subject, as was done in \cite{khodabakhshi2017, Islam2018, haider2022, Sen2021}, is more in alignment with the clinical setting. Accordingly, channel fusion should also be performed if the measurements are multi-channel. In this study, it was automatically performed by the VAR model as in \cite{Sen2021}, or by the CNN when it was using the spectro-temporal representations (MFCC, log-mel spectrogram and log-spectrogram) as input. Channel fusion was also done in \cite{Islam2018} and \cite{Messner2020}. 

Various augmentation scenarios based on noise infusion and mixup were also applied to the entire dataset in this study to test whether they would improve the results. Augmentation was also done in \cite{Srivastava2021}, although only one of the two classes was augmented in order to achieve comparable sample sizes. 

Note that the five studies that performed asthma vs. COPD classification \cite{khodabakhshi2017, Islam2018, haider2022, koshta2023, Sen2021} used different datasets. This study used the same local dataset as \cite{Sen2021} (i) to be able to compare the results, (ii) because even in the ICBHI database, the largest open pulmonary sounds database to date, there is only one subject with asthma (against 64 COPD samples), and the flow sub-phases are not annotated. Therefore, the performance scores reported here should only be compared across the different scenarios tested within this study, and the highest score should only be compared against that of the study in \cite{Sen2021}, as it is the only comparable study due to the common dataset and diagnostic purpose.

The rest of the paper is organized as follows. Section \ref{sec:data} describes the data used in this research. Section \ref{sec:Methods} provides an overview of the algorithms used, along with the basic theoretical information. Section \ref{sec:ExperimentalDesign} describes the design of the experiments and introduces the parameter selections to be compared, with detailed explanations. Section \ref{sec:ResultsDiscussions} reports the results along with the discussions. Finally, Section \ref{sec:Conclusion} draws conclusions and addresses possible future studies.

\section{Data}
\label{sec:data}

Pulmonary sounds data of $50$ volunteers were used in the study, where $30$ of them were diagnosed with asthma and $20$ of them with COPD, at Yedikule Chest Disease and Thoracic Surgery Education and Research Hospital, Istanbul, Turkey. 

The data were recorded via the 14-channel pulmonary sound data acquisition and processing system \cite{Sen2005} that was designed and implemented at the Bogazici University Lung Acoustics Laboratory. The system acquired 14-channel pulmonary sounds on the posterior chest wall and also measured the flow signal (i.e., flow rate of the inhaled and exhaled air in liters per second) simultaneously for synchronization of the sounds over the breathing cycle. During the recording sessions, the subjects were sitting upright and wearing a nose clip while breathing spontaneously through the mouthpiece of the flow meter. The data were sampled at a rate of 9600 samples per second. An acquisition session lasted 15 seconds, therefore each recording covers multiple breathing cycles (full-cycles) from the patient. Some useful details about the dataset including the full-cycle counts and varying full-cycle durations are given in Table \ref{tab:dataset_info}. Further details about the system hardware, data acquisition protocol, clinical information about the sample groups, and microphone locations on the chest wall can be found in \cite{Sen2021} and \cite{Sen2005}. The study had the approval of the Second Ethical Committee of Clinical Researches of Istanbul (in compliance with the Declaration of Helsinki) and an informed consent was taken from each subject before recording. 

\begin{table}[t]
\centering
\caption{Some details about the dataset \label{tab:dataset_info}}
\begin{tabular}{|l|c|c|c|} \hline
& \textbf{Asthma} & \textbf{COPD} & \textbf{Overall} \\ \hline
Number of patients & 30 & 20 & 50\\ \hline
Number of full-cycles & 139 & 89 & 228\\ \hline
Longest full-cycle duration (s) & 5.06 &  6.15 &  6.15\\ \hline
Shortest full-cycle duration (s) & 1.31 & 1.70 & 1.31\\ \hline
Average full-cycle duration (s) & 2.56 & 2.63 & 2.59\\ \hline
\end{tabular}
\end{table}

To prepare the data for the calculation of the two-dimensional signal representations, the start and end indices of the inspiration and expiration periods, as well as those of the sub-phases, were determined using the flow signal. As described in \cite{Sen2021}, the early and late phases of inspiration (or expiration) correspond to $30$ percent of the total volume of air inhaled (or exhaled), while the mid phase corresponds to the remaining $40$ percent. Using these indices, the pulmonary sounds were divided into full-cycle or sub-phase portions, depending on the experiment to be performed. Then from these sound portions, the two-dimensional representations described in Section \ref{Subsec:Two-dimInpRep} were computed.  

\section{Methods}
\label{sec:Methods}

\subsection{Signal Representations}
\label{Subsec:Two-dimInpRep}

The two-dimensional representations calculated from pulmonary sounds are:
\begin{itemize}
    \item Spectro-temporal representations to summarize the time-frequency behavior of the signal per channel, namely:
    \begin{itemize}
        \item Log spectrogram,
        \item Log-mel spectrogram,
        \item Mel-frequency cepstral coefficients (MFCC), and
    \end{itemize}
    \item Vector auto-regressive (VAR) model coefficient matrices to summarize the spatio-temporal relationships between the channels.
\end{itemize}

The spectro-temporal representations were obtained either for the full-cycles (yielding one matrix per full-cycle), or, for the sub-phases (yielding one matrix per sub-phase, i.e., a set of six matrices per full-cycle), depending on the scenario being tested. The VAR model matrices, on the other hand, were obtained for shorter signal segments (such that many segments fit in one sub-phase) due to the stationarity requirement of the model. In that case, one full-cycle was represented by multiple sets of six matrices (not by a single matrix or one set of six matrices).

The theoretical information on the computation of these representations are given below. On the other hand, the hyperparameter values to be compared through the experiments (e.g., the time and frequency resolution for the spectro-temporal representations) will be reported in Section \ref{sec:ExperimentalDesign}, along with the rationale behind their selection.

\subsubsection{Spectro-temporal Representations} \label{Sec:SpectroTemporalRep}

\textbf{Log spectrogram} was computed by taking the logarithm (to base ten) of the power spectrogram (the magnitude square of the STFT) then multiplying the result by ten. In the computation of the STFT, $L_{FFT}$-sample sliding windows of type Hanning were used with an overlap of 50$\%$ and the number of FFT points is set to $N_{FFT}$ ($N_{FFT}\geq L_{FFT}$). 

The number of frequency points (or, spectral resolution) in the STFT (hence the power spectrogram, and hence the log spectrogram) is
\begin{equation}
N_f=N_{FFT}/2+1
\label{Eq:freqRes}
\end{equation}
since only the non-negative frequencies are considered. This value is the total number of points covering the range from zero to the maximum frequency (half the sampling rate according to Nyquist's theorem), so it must be chosen wisely to provide sufficient resolution to represent important sound events. The number of time points (or, temporal resolution) $N_t$ can be calculated using 
\begin{equation}
N_t = \frac{2L_s}{L_{FFT}}-1
\label{Eq:timeRes}
\end{equation}
where $L_s$ is the total length of the signal (in samples) to be represented by the STFT.

\textbf{Log-mel spectrogram} was obtained by first computing the power spectrogram, then applying the mel filter bank \cite{Davis1980}, and finally multiplying the logarithm (to base ten) of the result by ten. The number of mel bands (hence, the spectral resolution for this representation) will be denoted by $N_{Mel}$ hereafter. Logarithmic magnitude scale was preferred over linear for both the spectrogram and mel spectrogram here, since the dynamic range it provides is more useful to describe the frequency characteristics of the signal.

\textbf{Mel-frequency cepstral coefficients (MFCC)} were computed by taking the Discrete Cosine Transform (DCT) of the log-mel spectrogram \cite{Logan2000}. DCT of Type-II with orthogonal normalization was used here. All of the coefficients were kept after the DCT, i.e., the number of MFCC was equal to the number of triangular mel filters. Therefore, MFCC matrices had the same spectral and temporal resolutions as log-mel spectrograms.

As stated above, the spectro-temporal representations were obtained either for the full-cycles or for the sub-phases. Varying durations of full-cycles or sub-phases result in unequal time resolutions, and this is one of the challenges in pulmonary sound classification using this type of input representations, since fixed input dimensionality is required to train the models. The frequency resolution of the spectro-temporal representation is determined by $N_{FFT}$, which is set to an appropriate value considering the longest $L_{FFT}$ in the experiment and the maximum frequency in the signal (half the sampling rate by Nyquist theorem); hence, it remains fixed regardless of varying signal durations. To fix the temporal resolution, on the other hand, two approaches can be (and were, in this study) adopted: 

\begin{itemize}

\item \textit{Trimming/padding} the signal to fix the time duration before computing the spectro-temporal representations. Trimming refers to truncating the signal, while padding (zero-padding) refers to appending zeros to it. In the literature, previous studies set a constant time duration by either trimming or padding the signals, thereby fixing the dimensions of the representations extracted from the signal \cite{Srivastava2021, wang2025, Messner2020, koike2021}. Considering the high variation in full-cycle lengths, the trimming/padding solution may lead to notable defects in the models, as trimming causes losses in information for the cycles longer than the selected duration, and padding generates excessive incorrect information. The only way to avoid this, while at the same time equalising the matrix dimensions,  is to apply a long window to long signals and a short window to short ones, i.e., by making the window length adaptive.

\item \textit{Adaptive-length windowing}, i.e., adjusting the lengths of the STFT windows adaptively to fix the resulting temporal resolution. For a given temporal resolution $N_t$, the length of the adaptive window to be applied, $L_{FFT}$, is calculated from (\ref{Eq:timeRes}) as: 
\begin{equation}
    L_{FFT} = \frac{2 L_s}{N_t + 1}
    \label{eq:findLFFT}
\end{equation}
In this case, the representations convey information from the entire signal in $N_t$ points, regardless of its duration. The concern here would be about reflecting the frequency content accurately when the window is considerably short, i.e., when $L_{FFT} \ll N_{FFT}$, $N_{FFT}$ being fixed based on the longest such window in the overall data for a given selection of $N_t$.

Both approaches have their own disadvantages, however, these are expected to be reflected in their respective errors and the comparison after the experiments will determine which is more useful.

\end{itemize}

\subsubsection{VAR Model}
The VAR model was assessed to be very successful in the differential diagnosis of asthma and COPD in the previous study \cite{Sen2021}. Based on the studies in \cite{Sen2021} and \cite{Sen2013}, where further details can be found, the order of the model and the sample size (the length of the short signal segments that the model is fitted on, in samples) were set to two and $250$, respectively, and 50\% overlapping was applied to avoid loss of information around the edges. The size of a VAR matrix is 14 (the number of channels) by 28 (14 times two, the model order), and the elements are the model coefficients that describe the spatio-temporal relationships between the channels (microphones). 

First, a signal segment of $250$ samples (approximately $26$ ms) cover a much shorter time interval than the duration of a typical full-cycle (say, $1.31$ s, the shortest one in the set, see Table \ref{tab:dataset_info}) or even sub-phase (say, $1.31 / 6 = 218$ ms roughly). Second, there is overlap between the two successive segments. Due to these two reasons, multiple models fitted in, i.e., multiple matrices were obtained from, each one of the sub-phases. Therefore, as introduced above, one full-cycle could not be represented by a single matrix or by a single set of six matrices, which was the case with the spectro-temporal representations. Instead, those VAR matrices with the same order index within the six sub-phases were brought together to form a set of six matrices, and one full-cycle was represented by many such sets. Note that a set was not formed at all if all six peers were not available due to unequal sub-phase lengths.

To summarize the differences between the two representation types, namely, a spectro-temporal representation and a VAR matrix, (i) the former describes the time-frequency behavior, whereas the latter depicts the spatio-temporal relationships, (ii) the former refers to one channel each (to be combined later), whereas the latter inherently holds the 14-channel information, (iii) the former inherently involves time variation, whereas the latter gives a static picture with no inherent time variation (the course of time can only be observed by sequencing the VAR matrices one after the other).

\subsection{Preparing the Representations as Inputs}
\label{Subsec:PreparingRepAsInputs}
Although the spectro-temporal representation obtained for the full-cycle already contains the overall information, computation for the six sub-phases separately was also performed (i) for a more accurate comparison with the VAR model implementation, where the matrices are sub-phase specific, (ii) for a more controlled and effective use of timing information by the classifier. In this case, regardless of the representation type, combining (fusing) the sub-phases either in the representation space or in the feature space were considered for investigation. Note that the words \textit{representation} and \textit{feature} are used cautiously throughout the manuscript, referring to the stages before and after a dedicated feature extractor network that learns from the two-dimensional representations, respectively.

Accordingly, various scenarios were designed to prepare the inputs to train the classifier models, according to the strategy adopted for fusing the information provided by the sub-phases:
\begin{itemize}

    \item \textit{Full-cycle representation}: Only valid with the spectro-temporal representations. The representation matrix was computed for the overall full-cycle (one matrix per full-cycle), resulting in input dimensions of $N_t\times N_f\times14$, where $14$ is the number of channels (microphones), $N_f$ is the spectral dimension as defined in (\ref{Eq:freqRes}), and $N_t$ is the temporal dimension as defined in (\ref{Eq:timeRes}) and is associated with one full-cycle here. Note that the sound signal was taken from the first index of the inspiration till the last index of the expiration that follows. That is, although the sub-phase indices associated with the particular full-cycle were also known, they were ignored with this type of representation.
    
    \item \textit{Joined representation}: The representations obtained from the six sub-phases of the same full-cycle were concatenated to form a joined representation (fusing was done in the representation space). With the spectro-temporal representations, this resulted in input dimensions of $(6*N_t)\times N_f\times14$, where $N_t$ is associated with one sub-phase now and was approximately set to one sixth of the value used for the full-cycle representations, to keep more or less the same overall temporal dimension for a full-cycle. With the VAR model, the inputs had dimensions of $(6*14) \times 28 \times 1$, and while concatenating, the matrices with the same order index within the sub-phase were brought together. 
    
    \item \textit{Separate representation}: Sub-phases were represented individually as in the joined representation scenario, however they were not concatenated directly in the representation space. Instead, the six individual representations were fed into six separate feature extractor networks, and then fusing was done in the feature space. This enables training multiple networks, each being dedicated to one of the six sub-phases. Input dimensions were $N_t \times N_f \times 14$ with the spectro-temporal representations and $14 \times 28 \times 1$ with the VAR model, respectively. 
    
\end{itemize}

\subsection{Deep Learning Algorithms}
\label{Subsec:DeepLearningAlg}

\begin{figure}[t]
    \centering
    \begin{subfigure}[]{.8\textwidth}
    \includegraphics[width=\textwidth]{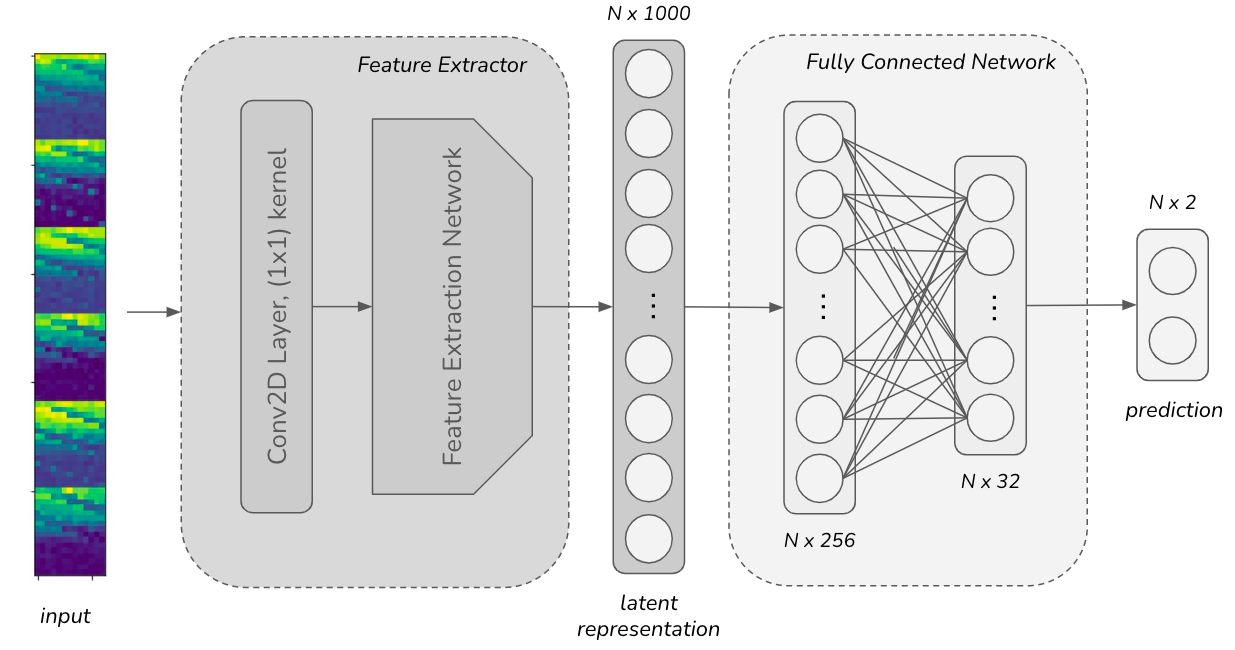}
    \caption{Single-modal Network.}
    \end{subfigure}
    
    \begin{subfigure}[]{.8\textwidth}
    \includegraphics[width=\textwidth]{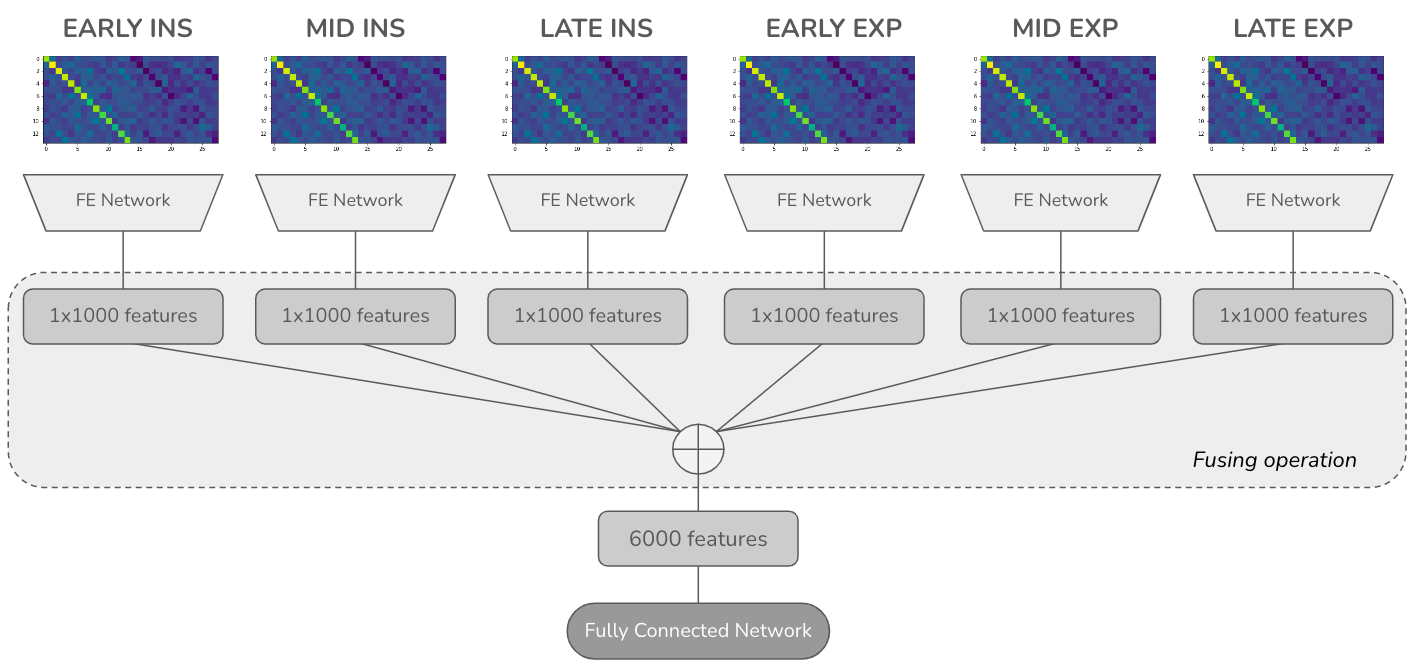}
    \caption{Multi-modal Network.}
    \end{subfigure}
    \caption{Overall structure of the networks. *FE: Feature extractor \label{fig:overall_structure}}
\end{figure}

Just as the different representation designs are named specially for better differentiation, so should the models that accept them as inputs. Accordingly, the classifier with a single feature extractor applied on the overall full-cycle was called \textit{single-modal}, while the classifier with the six feature extractors, each being dedicated to one sub-phase, was called \textit{multi-modal}. In this case, full-cycle representations and joined representations were fed into single-modal feature extractors, while separate representations were fed into multi-modal feature extractors. Then, the extracted features were passed through the classification network to perform the prediction. Figure \ref{fig:overall_structure} exhibits the overall structure of the networks.

The \textit{single-modal} network comprised a single feature extractor module aiming to obtain the latent vector from the given full-cycle representation or joined representation. With the full-cycle representation, the overall full-cycle behavior was contained in a single input. However, previous work has shown that certain sub-phases contain more important information than the others in terms of distinguishing asthma and COPD \cite{Sen2021}. Therefore, with the joined representation, the focus of the network on the sub-phases was intuitively enhanced. 

The \textit{multi-modal} network consisted of six individual feature extractor modules, each being dedicated to one sub-phase of the given full-cycle. The isolation of each unit permits the network to focus on each sub-phase respectively and learn the salient features separately. Before the classification network, the latent representations according to the sub-phases provided by individual modules were fused to present the feature representation of the signal, the details of which will be provided in the subsequent paragraphs.

Convolutional Neural Network (CNN) and Temporal Convolutional Network (TCN) architectures were used for feature extraction in both the single-modal and multi-modal scenarios, and a specialized version of Recurrent Neural Networks (RNN), namely, Gated Recurrent Units (GRU) architecture was used for feature fusion in the multi modal scenario. 

CNNs have become the focus of many studies in several fields owing to their ability in capturing spatially salient features, such as computer vision \cite{farrajota2019,hossain2019}, psychology \cite{jha2023}, or biomedical engineering \cite{Srivastava2021, zhang2024, shehab2024, jung2021, kim2021, asatani2021, zulfiqar2021, wang2024, kim2025, iqbal2021}. CNN architecture enables learning the latent representations of the given information by utilizing convolution responses of trainable weights on the sliding kernels through the data. In particular, using multi-dimensional kernels permits the models to capture information from multiple domains. Taking signal processing as an example, the usage of spectro-temporal representations of the signal, like spectrograms, elevates the capabilities of CNNs by providing short-time analysis of the signal.

To this extent, a variety of large-scale CNN architectures, namely ResNet \cite{he2016}, Wide-ResNet \cite{zagoruyko2016}, DenseNet \cite{huang2017}, and VGG \cite{simonyan2014}, were employed in the feature extraction modules in this study. Although large-scale models provide deep networks, they suffer from the vanishing gradient problem that constrains the training of earlier layers in the network. To alleviate this problem, residual networks like ResNet utilize residual connections aggregating intermediate outcomes to the posterior layers of the network. Nevertheless, they are still prone to overfitting, especially in small dataset sizes \cite{pavlitskaya2022}. Therefore, a custom shallow CNN architecture was also designed, which benefits from having fewer layers and is intuitively less susceptible to the detrimental effects of data scarcity. The details about the shallow CNN model are presented in Table \ref{tab:shallow_network}.

An alternative convolutional architecture, namely TCN, proposed by Bai \textit{et al.} has shown superior performance over recurrent models on sound analysis tasks in sequence modelling \cite{bai2018}. The TCN architecture comprises multiple sequentially connected Residual Convolutional Blocks, which utilize 1-dimensional convolutional layers with dilated kernels. The performance of the TCN architecture on pulmonary sounds has also been explored in this study along with the above-mentioned CNN models. 

Although RNNs are suitable for sequential information, the computational cost exceptionally increases as the sequence length increases. Since the signal sequence length for raw sounds is high considering the sampling rate, processing the raw signal by RNN is impractical. Therefore, in this study, the RNN architecture was used with the lower-dimensional representations of the extracted features, for their fusing in the multi-modal network (see Figure \ref{fig:overall_structure}).

\begin{table}[t]
\centering
\caption{Overall architecture of Shallow CNN model\label{tab:shallow_network}}
\begin{tabular}{|c|lcc|}
\hline
\textbf{Block} & \multicolumn{1}{l|}{\textbf{Layer}} & \multicolumn{1}{c|}{\textbf{Unit Size}} & \textbf{Kernel Size} \\ \hline
\multirow{5}{*}{1} & \multicolumn{1}{l|}{Conv2D} & \multicolumn{1}{c|}{256} & 1x1 \\ \cline{2-4} 
 & \multicolumn{1}{l|}{Conv2D} & \multicolumn{1}{c|}{128} & 3x3 \\ \cline{2-4} 
 & \multicolumn{1}{l|}{Conv2D} & \multicolumn{1}{c|}{64} & 3x3 \\ \cline{2-4} 
 & \multicolumn{3}{c|}{ReLU} \\ \cline{2-4} 
 & \multicolumn{3}{c|}{BatchNormalization} \\ \hline
\multirow{4}{*}{2} & \multicolumn{1}{c|}{Conv2D} & \multicolumn{1}{c|}{32} & \multicolumn{1}{c|}{5x5} \\ \cline{2-4} 
 & \multicolumn{3}{c|}{ReLU} \\ \cline{2-4} 
 & \multicolumn{3}{c|}{BatchNormalization} \\ \cline{2-4} 
 & \multicolumn{3}{c|}{MaxPooling} \\ \hline
\end{tabular}
\end{table}

In this study, three strategies were adopted for fusing the sub-phase-specific features in the multi-modal scenario: 

\begin{itemize}

\item \textit{Concatenation:} direct concatenation of the features extracted for the six sub-phases. The studies in multi-modal structures showed that straightforward procedures such as concatenation exhibit better performance compared to complex approaches \cite{siriwardhana2020}. Therefore, the baseline is set with concatenation fusing. 

\item Fusing by \textit{GRU:} Since the extracted features carry the temporal order property of sub-phases, RNNs were experimented, advancing the baseline fusing strategy. In this study, GRU architecture, a specialized version of RNN, was employed for ensembling the feature vectors of the sub-phases, upon examination of their temporal relations. 

\item Fusing by \textit{GRU+Attention:} Considering the fact that some of the sub-phases have more prominent effect on the diagnosis, a Bahdanau attention mechanism \cite{bahdanau2014} has also been utilized along with GRU, for amplifying the salient features coming from particular sub-phases. 

\end{itemize}

To evaluate the network's decision out of the extracted (or extracted then fused) features, a fully-connected neural network is employed. Experimental trials have revealed that the use of the \textit{tanh} activation function in the output layer of the network provides more significant gradient information than the \textit{softmax} function while training and leads to an overall improvement in network performance. During inference, the output was redistributed with the softmax function to provide the associated probabilities.

\subsection{Data Augmentation}
\label{Subsec:DataAug}
Useful data is usually scarce in clinical data processing studies due to factors such as missing annotation, outside noise, or limited amount of volunteers who meet the eligibility criteria. On the other hand, deep learning networks are prone to under or overfitting when the data size is limited. Two types of data augmentation methods were tested in this study, with an attempt to assess their effectiveness in this particular situation.

Initially, the sound samples were augmented by random noise infusion. Accordingly, white noise is added to each channel of the pulmonary sounds. Considering that high noise intensity may deceive the models, the added noise is limited such that the generated sample would not fall below a certain signal-to-noise ratio (SNR). 

In addition to noise infusion, mixup augmentation \cite{zhang2017} was employed during the training of the models to sustain stable training. In mixup augmentation, the generated input $\tilde{x}$ and the ground truth $\tilde{y}$ were calculated as
\begin{equation} \label{eq:mixup}
    \begin{split}
        \tilde{x} = \lambda x_i + (1 - \lambda) x_j \\ 
        \tilde{y} = \lambda y_i + (1 - \lambda) y_j    
    \end{split}
\end{equation}
where $x_i$ and $x_j$ refer to input $i$ and $j$, $y_i$ and $y_j$ refer to the corresponding ground truths, and $\lambda$ is a random value sampled from the $\mathrm{Beta}$ distribution. Generating a synthetic combination of inputs and ground truths compels models to minimize error for each class more precisely. In this way, models are encouraged to be more confident in their decision to distinguish between the two diseases.

Both methods were investigated in various combinations, such that various SNR values were tested in different combinations and each combination was tested with and without mixup augmentation. The options will be described in Section \ref{sec:ExperimentalDesign}.

\section{Experimental Design}
\label{sec:ExperimentalDesign}
The experiments were conducted in leave-pair-out 10-fold cross-validation configuration. Due to the limitation in data size, three asthma and two COPD patients were reserved for testing in each iteration. Similarly, one asthma and one COPD patient, different from those in the test set, were used for validation. The rest of the patients were employed for training. Network weights were randomly initialized at each iteration of the cross-validation. Since it may lead to inconsistencies in the results, especially in a low-resourced dataset, the experiments were repeated thrice to provide a robust assessment. All the results reported here were obtained in this way, i.e., over 30 independent experiments with full coverage of all the patients in the dataset.

In all the experiments, the models were set to extract 1000-dimensional feature vectors from each sub-phase. In TCN architecture, two consecutive temporal convolutional blocks with unit size of 64 were used. The networks were trained using Adam optimizer with a learning rate of $10^{-4}$. The cross-entropy loss function was used to assess the error in the predictions. In terms of early stopping, the training was halted when any improvement in the validation loss was not observed in consecutive 20 epochs. The weights achieving the best F1-score (positive class: COPD) on the validation set during the training were stored and used for the performance evaluation on the test set. 

The \textbf{first} experiment was designed to test the model architectures to be used for feature extraction given the spectral representations. The shallow CNN network (of Table \ref{tab:shallow_network}), ResNet18, ResNet50, Wide ResNet50, VGG11, DenseNet121, and TCN were tested. Among the selections $N_{Mel}=13,26,39$ for MFCC and log-mel spectrogram, the smallest two were not included since the resolution on the frequency axis was insufficient for the depth of some of the models. The lowest and highest temporal resolution selections from both the trimming/padding and the adaptive-length windowing scenarios were included, as the remaining would be time consuming without significant benefit.

The \textbf{second} experiment was designed to test whether the trimming/pad-ding scenario or the adaptive-length windowing scenario should be preferred when obtaining the full-cycle spectro-temporal representations. Upon observing the average full-cycle durations in Table \ref{tab:dataset_info} in Section \ref{sec:data}, the fixed time duration was chosen to be 3 seconds ($L_s=28800$ samples) for the trimming/padding scenario. Any full-cycle (actually, the synchronous pulmonary sound signal) longer than this was trimmed from the end, and any full-cycle shorter than this was zero-padded at the end, before computing the STFT. The number of FFT points was set to $N_{FFT} = 1024$ in order to obtain a spectral resolution ($512$ points) sufficient to represent $0$ to $4800$ Hz (the maximum frequency due to the Nyquist Theorem, with the sampling rate of 9600 Hz) without increasing the feature size too much along the vertical axis. The selections for the window length were $L_{FFT} = 1024, 512, 256, 128$. Longer windows cannot be applied as $L_{FFT} \le N_{FFT}$ and shorter windows along with $50\%$ overlapping increase temporal resolution unnecessarily (increase the feature size too much along the horizontal axis) while failing to capture meaningful acoustic waveforms in pulmonary sounds (e.g., $L_{FFT}=64$ corresponds to 6.7 ms, and this is shorter than a typical course crackle \cite{sovijarvi2000p597, ATS}). Table \ref{tab:winlengths} lists the resulting temporal resolution $N_t$ (second column) per selection of window length $L_{FFT}$ (first column), calculated using (\ref{Eq:timeRes}). For the adaptive-length windowing  scenario, on the other hand, the resulting $N_t$ (third column) was first decided, then $L_{FFT}$ was calculated accordingly. The temporal resolution selections in the adaptive-length windowing scenario, namely $N_t = 128, 256, 512$, were specifically chosen to be the closest power-of-two integers to those of the trimming/padding scenario (compare the third column with the second), then for every given full-cycle of length $L_s$, window length $L_{FFT}$ was calculated using (\ref{eq:findLFFT}). Taking $N_t = 64$ and equating $L_s$ to the length of the longest cycle ($6.15$ seconds, i.e., $59040$ samples, see Table \ref{tab:dataset_info}) yields $L_{FFT} = 1817$ approximately. This is greater than the fixed value of $N_{FFT} = 1024$, therefore the selection of $N_t = 64$ has not been applied. The shallow CNN network and ResNet18 were used in this experiment because of the difficulty in training the deeper models with the small representation sizes to be tested.

\begin{table}[t]
\centering
\caption{Temporal resolutions in the trimming/padding and adaptive-length windowing scenarios of the second experiment.} \label{tab:winlengths}
\begin{tabular}{|c|c|c|} \hline
\textbf{trim/pad $L_{FFT}$} & \textbf{trim/pad $N_t$} & \textbf{adapt-win $N_t$} \\ \hline
1024    & 55    & 64*       \\ \hline
512     & 111   & 128      \\ \hline
256     & 224   & 256     \\ \hline
128     & 449   & 512     \\ \hline
\end{tabular}\\
\hspace{-5.5cm} * not applied
\end{table}

The \textbf{third} experiment was designed to test the optimal time resolution for the spectro-temporal representations of sub-phases to be used in the joined representation. As six such representations were to be concatenated along the time axis to represent a full-cycle, the optimal time resolution found in the second experiment (adaptive-length windowing scenario, $N_t = 256$) was divided by six (yielding approximately $43$), then the time resolution values to be tested were chosen to cover a range around this value. Consequently, the set of values experimented was (power-of-two integers as a convention without a mathematically significant reason) $N_t = 16, 32, 64, 128$. Given the selected value of $N_t$, $L_{FFT}$ was calculated using (\ref{eq:findLFFT}) where $L_s$ is the length of the sub-phase under consideration this time. Then, $N_{FFT}$ was set to the smallest power-of-two integer such that $N_{FFT} \geq L_{FFT}$. The VAR model was also included in this experiment for comparison. With the spectro-temporal representations, where a full-cycle was represented by one set of six matrices, note that the success score was calculated over the full-cycles (i.e., the proportion of correctly classified full-cycles give the accuracy). However, with the VAR model, where a full-cycle was represented by multiple sets of six matrices, the success score was calculated over those sets (i.e., the proportion of correctly classified sets gave the accuracy).

Having decided on the optimal time resolution for the spectro-temporal representations, the \textbf{fourth} experiment was carried out to investigate all possible strategies for combining the diagnostic contributions of the sub-phases: (i) single-modal perspective (with full-cycle and joined representations), (ii) multi-modal perspective (with separate representations) where features were (a) concatenated, or, fused by using (b) GRU, or (c) GRU+Attention. In GRU experiments, a single-layer bidirectional GRU model with 256 hidden dimensions was employed to analyze the temporal relation between the sub-phases. In GRU+Attention experiments, the output states of the GRU layer were passed through an attention layer in which the size of the attention dimension is set to 128. 

The \textbf{fifth} experiment was conducted to investigate the augmentation methods. Adding white noise with SNR values of 5, 10, 15 and 20 dB were tried in various combinations (i.e., 5 dB alone, 5 and 10 dB, and all four together). For the mixup augmentation, the parameters of the $\mathrm{Beta}$ distribution (from which $\lambda$ in (\ref{eq:mixup}) was sampled), i.e., $\alpha$ and $\beta$, were both set to 0.2. Noise infusion SNR value combinations were tested both with and without mixup augmentation.

Finally, the optimal classifiers pointed out by the five experiments were analyzed in terms of their \textbf{subject-based} diagnostic capabilities. For this purpose, subject-based decisions were obtained through majority voting over the full-cycle decisions. Alternatively, a strategy based on the Bayesian decision rule was tested, in which the network output was used as the likelihood, the proportion of full-cycles observed was used as the prior, and the decision was made by maximizing the calculated posterior probability.

Throughout the experiments, the F1-score (positive class: COPD) was used as the criterion for comparisons, and the choice with the highest F1-score was continued to the next experiment. Specificity, sensitivity and accuracy were also calculated to evaluate subject-based performances. Finally, classifier error distributions were compared using the unpaired t-test to see if one scenario had a statistically significant advantage over the other in terms of diagnostic success.

\section{Results and Discussions}
\label{sec:ResultsDiscussions}

The results of the experiments are reported below in order.  Please note that the verbs like \textit{to outperform} or \textit{to be better} do not always imply that the associated error distribution to be smaller in the mean with statistical significance ($p<0.05$). Since three repetitions of 10-fold cross-validation were conducted per experiment, yielding a total of 30 numerical values, a higher average score (despite not being statistically significant according to the t-test) will still be interpreted as meaningful, particularly as a recommended choice when selecting between two alternatives.

The results (F1-scores) of the \textbf{first} experiment (testing the model architectures) are given in Table \ref{tab:exp1}. The shallow CNN network (0.7565) outperformed all other model architectures tested, followed by ResNet18 and DenseNet121 (0.7318 both). Observing that DenseNet121 outperforms Res-Net18 for only one of the four options (the rigthmost column of the four) in the upper half of the table (where the highest scores lie), it can be concluded that the shallow model and ResNet18 were the two models selected by the first experiment. The success of the simpler models can be attributed to the adequacy of lower complexity, as well as to the size of the dataset.

%%%%%%%%%%%%%%%%%%%% 111111111111111111111111111111 %%%%%%%%%%%%%%%%%%%%%%%%%%%%%%
\begin{table}[b]
\caption{F1-scores of the first experiment (testing the model architectures).} \label{tab:exp1}
\centering
\begin{tabular}{cr|cc|cc|}
\cline{3-6}
\multicolumn{1}{l}{}                                         & \multicolumn{1}{l|}{}   & \multicolumn{2}{c|}{\textbf{trim/pad $L_{FFT}$}}                                               & \multicolumn{2}{c|}{\textbf{adapt-win $N_t$}}                                                                              \\ \cline{3-6} 
 &  & \multicolumn{1}{c|}{\textbf{1024}}              & \textbf{128}               & \multicolumn{1}{c|}{\textbf{128}}                                               & \textbf{512}               \\ \hline
\multicolumn{1}{|c|}{}                                       & \textbf{Shallow}        & \multicolumn{1}{c|}{\cellcolor[HTML]{F4F4F4}0.7036} & \cellcolor[HTML]{919191}0.1902 & \multicolumn{1}{c|}{\cellcolor[HTML]{FFFFFF}{\color[HTML]{C00000} \textbf{0.7565}}} & \cellcolor[HTML]{878787}0.1385 \\ \cline{2-6} 
\multicolumn{1}{|c|}{}                                       & \textbf{ResNet18}       & \multicolumn{1}{c|}{\cellcolor[HTML]{E0E0E0}0.5980} & \cellcolor[HTML]{F5F5F5}0.7086 & \multicolumn{1}{c|}{\cellcolor[HTML]{FAFAFA}0.7318}                                 & \cellcolor[HTML]{EDEDED}0.6684 \\ \cline{2-6} 
\multicolumn{1}{|c|}{}                                       & \textbf{ResNet50}       & \multicolumn{1}{c|}{\cellcolor[HTML]{DBDBDB}0.5731} & \cellcolor[HTML]{D9D9D9}0.5635 & \multicolumn{1}{c|}{\cellcolor[HTML]{DADADA}0.5685}                                 & \cellcolor[HTML]{E0E0E0}0.5979 \\ \cline{2-6} 
\multicolumn{1}{|c|}{}                                       & \textbf{Wide ResNet50}  & \multicolumn{1}{c|}{\cellcolor[HTML]{E4E4E4}0.6174} & \cellcolor[HTML]{F5F5F5}0.7079 & \multicolumn{1}{c|}{\cellcolor[HTML]{EFEFEF}0.6771}                                 & \cellcolor[HTML]{DADADA}0.5660 \\ \cline{2-6} 
\multicolumn{1}{|c|}{}                                       & \textbf{VGG11}          & \multicolumn{1}{c|}{\cellcolor[HTML]{878787}0.1417} & \cellcolor[HTML]{9A9A9A}0.2368 & \multicolumn{1}{c|}{\cellcolor[HTML]{A6A6A6}0.2970}                                 & \cellcolor[HTML]{BDBDBD}0.4197 \\ \cline{2-6} 
\multicolumn{1}{|c|}{}                                       & \textbf{DenseNet121}    & \multicolumn{1}{c|}{\cellcolor[HTML]{DEDEDE}0.5907} & \cellcolor[HTML]{E8E8E8}0.6397 & \multicolumn{1}{c|}{\cellcolor[HTML]{EAEAEA}0.6522}                                 & \cellcolor[HTML]{FAFAFA}0.7318 \\ \cline{2-6} 
\multicolumn{1}{|c|}{\multirow{-7}{*}{\textbf{mfcc-39}}}     & \textbf{TCN}            & \multicolumn{1}{c|}{\cellcolor[HTML]{808080}0.1005} & \cellcolor[HTML]{919191}0.1902 & \multicolumn{1}{c|}{\cellcolor[HTML]{A2A2A2}0.2795}                                 & \cellcolor[HTML]{ABABAB}0.3252 \\ \hline
\multicolumn{1}{|c|}{}                                       & \textbf{Shallow}        & \multicolumn{1}{c|}{\cellcolor[HTML]{D6D6D6}0.5457} & \cellcolor[HTML]{CCCCCC}0.4933 & \multicolumn{1}{c|}{\cellcolor[HTML]{DCDCDC}0.5769}                                 & \cellcolor[HTML]{DCDCDC}0.5790 \\ \cline{2-6} 
\multicolumn{1}{|c|}{}                                       & \textbf{ResNet18}       & \multicolumn{1}{c|}{\cellcolor[HTML]{D8D8D8}0.5600} & \cellcolor[HTML]{EDEDED}0.6672 & \multicolumn{1}{c|}{\cellcolor[HTML]{EAEAEA}0.6509}                                 & \cellcolor[HTML]{D2D2D2}0.5267 \\ \cline{2-6} 
\multicolumn{1}{|c|}{}                                       & \textbf{ResNet50}       & \multicolumn{1}{c|}{\cellcolor[HTML]{C8C8C8}0.4749} & \cellcolor[HTML]{E0E0E0}0.5971 & \multicolumn{1}{c|}{\cellcolor[HTML]{D4D4D4}0.5387}                                 & \cellcolor[HTML]{D6D6D6}0.5464 \\ \cline{2-6} 
\multicolumn{1}{|c|}{}                                       & \textbf{Wide ResNet50}  & \multicolumn{1}{c|}{\cellcolor[HTML]{DCDCDC}0.5761} & \cellcolor[HTML]{DBDBDB}0.5739 & \multicolumn{1}{c|}{\cellcolor[HTML]{DDDDDD}0.5845}                                 & \cellcolor[HTML]{E4E4E4}0.6221 \\ \cline{2-6} 
\multicolumn{1}{|c|}{}                                       & \textbf{VGG11}          & \multicolumn{1}{c|}{\cellcolor[HTML]{8F8F8F}0.1803} & \cellcolor[HTML]{A7A7A7}0.3069 & \multicolumn{1}{c|}{\cellcolor[HTML]{9A9A9A}0.2379}                                 & \cellcolor[HTML]{A5A5A5}0.2943 \\ \cline{2-6} 
\multicolumn{1}{|c|}{}                                       & \textbf{DenseNet121}    & \multicolumn{1}{c|}{\cellcolor[HTML]{EAEAEA}0.6482} & \cellcolor[HTML]{E7E7E7}0.6370 & \multicolumn{1}{c|}{\cellcolor[HTML]{E0E0E0}0.6003}                                 & \cellcolor[HTML]{E6E6E6}0.6292 \\ \cline{2-6} 
\multicolumn{1}{|c|}{\multirow{-7}{*}{\textbf{logmelsp-39}}} & \textbf{TCN}            & \multicolumn{1}{c|}{\cellcolor[HTML]{B0B0B0}0.3533} & \cellcolor[HTML]{B0B0B0}0.3517 & \multicolumn{1}{c|}{\cellcolor[HTML]{C0C0C0}0.4353}                                 & \cellcolor[HTML]{BFBFBF}0.4273 \\ \hline
\end{tabular}
\end{table}

The results of the \textbf{second} experiment (trimming/padding vs. adaptive-length windowing) are given in Table \ref{tab:exp2}. The scores reported here belong to the shallow network as it outperformed ResNet18. The maximum F1-score (0.8104) was achieved by MFCC with thirteen coefficients (MFCC-13) and when adaptive-length windowing  was applied for $N_t=256$. The corresponding trimming/padding ($L_{FFT}=256 \Rightarrow N_t=224$ by Table \ref{tab:winlengths}) F1-score is lower (0.7153), however not significantly (p=0.17). In general, when Table \ref{tab:exp2} is examined, it is observed that (i) MFCC \& adaptive-length windowing  combination gives the highest scores (brightest shades of the grayscale), and (ii) $N_t$ must be kept small as the number of MFCC is increased. When the same experiment was performed with ResNet18 (results not shown in a table for brevity), adaptive-length windowing  for $N_t=256$ was still the best choice (this time with MFCC-26, adaptive-length windowing : 0.7411\hspace{1mm}\textgreater\hspace{1mm}trimming/padding: 0.6736, however not significantly, p=0.33). Comparison between the best results of the two models did not show a statistically significant difference either (shallow CNN: 0.8104\hspace{1mm}\textgreater\hspace{1mm}ResNet18: 0.7411, p=0.28). The different drawbacks of trimming/padding and adaptive-length windowing  were outlined in Section \ref{Sec:SpectroTemporalRep}. The results of the second experiment show that the trimming/padding and adaptive-length windowing  scenarios are equally applicable; nevertheless, between the two, the latter is recommended based on its higher average success rates.

%%%%%%%%%%%%%%%%%%%% 222222222222222222222222222222 %%%%%%%%%%%%%%%%%%%%%%%%%%%%%%
\begin{table}[t] 
\caption{F1-scores of the second experiment (trimming/padding vs. adaptive-length windowing) with the shallow CNN.} \label{tab:exp2}
\centering
\begin{tabular}{r|cccc|ccc|}
\cline{2-8}
\multicolumn{1}{l|}{}                      & \multicolumn{4}{c|}{\textbf{trim/pad $L_{FFT}$}}                                                                                                                                                           & \multicolumn{3}{c|}{\textbf{adapt-win $N_t$}}                                                                                                                                   \\ \cline{2-8} 
\multicolumn{1}{l|}{}                      & \multicolumn{1}{c|}{\textbf{1024}}                  & \multicolumn{1}{c|}{\textbf{512}}                   & \multicolumn{1}{c|}{\textbf{256}}                   & \textbf{128}                   & \multicolumn{1}{c|}{\textbf{128}}                   & \multicolumn{1}{c|}{\textbf{256}}                                                   & \textbf{512}                   \\ \hline
\multicolumn{1}{|r|}{\textbf{mfcc-13}}     & \multicolumn{1}{c|}{\cellcolor[HTML]{E7E7E7}0.6816} & \multicolumn{1}{c|}{\cellcolor[HTML]{ECECEC}0.7079} & \multicolumn{1}{c|}{\cellcolor[HTML]{EDEDED}0.7153} & \cellcolor[HTML]{FBFBFB}0.7934 & \multicolumn{1}{c|}{\cellcolor[HTML]{FBFBFB}0.7882} & \multicolumn{1}{c|}{\cellcolor[HTML]{FFFFFF}{\color[HTML]{C00000} \textbf{0.8104}}} & \cellcolor[HTML]{F4F4F4}0.7520 \\ \hline
\multicolumn{1}{|r|}{\textbf{mfcc-26}}     & \multicolumn{1}{c|}{\cellcolor[HTML]{E7E7E7}0.6799} & \multicolumn{1}{c|}{\cellcolor[HTML]{F5F5F5}0.7576} & \multicolumn{1}{c|}{\cellcolor[HTML]{FDFDFD}0.8039} & \cellcolor[HTML]{C4C4C4}0.4855 & \multicolumn{1}{c|}{\cellcolor[HTML]{F5F5F5}0.7601} & \multicolumn{1}{c|}{\cellcolor[HTML]{FDFDFD}0.8016}                                 & \cellcolor[HTML]{858585}0.1330 \\ \hline
\multicolumn{1}{|r|}{\textbf{mfcc-39}}     & \multicolumn{1}{c|}{\cellcolor[HTML]{EBEBEB}0.7036} & \multicolumn{1}{c|}{\cellcolor[HTML]{EDEDED}0.7125} & \multicolumn{1}{c|}{\cellcolor[HTML]{CBCBCB}0.5231} & \cellcolor[HTML]{8F8F8F}0.1902 & \multicolumn{1}{c|}{\cellcolor[HTML]{F5F5F5}0.7565} & \multicolumn{1}{c|}{\cellcolor[HTML]{B1B1B1}0.3797}                                 & \cellcolor[HTML]{868686}0.1385 \\ \hline
\multicolumn{1}{|r|}{\textbf{logmelsp-13}} & \multicolumn{1}{c|}{\cellcolor[HTML]{DCDCDC}0.6178} & \multicolumn{1}{c|}{\cellcolor[HTML]{D4D4D4}0.5729} & \multicolumn{1}{c|}{\cellcolor[HTML]{E7E7E7}0.6777} & \cellcolor[HTML]{DDDDDD}0.6232 & \multicolumn{1}{c|}{\cellcolor[HTML]{DDDDDD}0.6272} & \multicolumn{1}{c|}{\cellcolor[HTML]{D9D9D9}0.6047}                                 & \cellcolor[HTML]{D3D3D3}0.5700 \\ \hline
\multicolumn{1}{|r|}{\textbf{logmelsp-26}} & \multicolumn{1}{c|}{\cellcolor[HTML]{D9D9D9}0.5997} & \multicolumn{1}{c|}{\cellcolor[HTML]{D7D7D7}0.5928} & \multicolumn{1}{c|}{\cellcolor[HTML]{D2D2D2}0.5618} & \cellcolor[HTML]{D9D9D9}0.6037 & \multicolumn{1}{c|}{\cellcolor[HTML]{CBCBCB}0.5271} & \multicolumn{1}{c|}{\cellcolor[HTML]{D6D6D6}0.5874}                                 & \cellcolor[HTML]{D0D0D0}0.5510 \\ \hline
\multicolumn{1}{|r|}{\textbf{logmelsp-39}} & \multicolumn{1}{c|}{\cellcolor[HTML]{CFCFCF}0.5457} & \multicolumn{1}{c|}{\cellcolor[HTML]{D0D0D0}0.5524} & \multicolumn{1}{c|}{\cellcolor[HTML]{D4D4D4}0.5721} & \cellcolor[HTML]{C5C5C5}0.4933 & \multicolumn{1}{c|}{\cellcolor[HTML]{D4D4D4}0.5769} & \multicolumn{1}{c|}{\cellcolor[HTML]{D1D1D1}0.5572}                                 & \cellcolor[HTML]{D5D5D5}0.5790 \\ \hline
\multicolumn{1}{|r|}{\textbf{logsp}}       & \multicolumn{1}{c|}{\cellcolor[HTML]{000000}-}      & \multicolumn{1}{c|}{\cellcolor[HTML]{000000}-}      & \multicolumn{1}{c|}{\cellcolor[HTML]{000000}-}      & \cellcolor[HTML]{000000}-      & \multicolumn{1}{c|}{\cellcolor[HTML]{B7B7B7}0.4144} & \multicolumn{1}{c|}{\cellcolor[HTML]{808080}0.1052}                                 & \cellcolor[HTML]{000000}-      \\
\hline
\end{tabular}
\end{table}
 
The results of the \textbf{third} experiment (testing the sub-phase temporal resolutions) are given in Table \ref{tab:exp3}. MFCC-13 achieved the highest score, while VAR could not attain this level despite being competitive (VAR: 0.7583\hspace{1mm}\textless\hspace{1mm} MFCC-13: 0.8108, however not significantly, p=0.29). In the table, the numbers given in parentheses are $N_{FFT}$ values used for the corresponding $N_t$ selection. As the adaptive window length was increased for smaller $N_t$, the default selection $N_{FFT}=1024$ was insufficient to satisfy $N_{FFT} \geq L_{FFT}$. Therefore, $N_{FFT}$ had to be increased first to $N_{FFT}=2048$ for $N_t=32$, then to $N_{FFT}=4096$ for $N_t=16$. Dividing $N_t=256$ (selection of the second experiment) by six (number of sub-phases) gives 42.7 (hypothetical $N_t$ per sub-phase in this case). Among the two closest options, namely $N_t=32$ and $N_t=64$, the greater achieved the best score. To fine-tune for the final implementation, the grid search can be expanded by including additional options around $N_t=64$. The success with the joined representation (0.8108, Table \ref{tab:exp3}) was almost identical (p=0.99) to that of the full-cycle representation (0.8104, Table \ref{tab:exp2}). ResNet18 (results not shown in a table for brevity) also yielded the best performance with MFCC-13 and $N_t=64$ (0.7740). This is better than its full-cycle representation performance with MFCC-13 and $N_t=256$ (0.6588), however not significantly (p=0.59). VAR failed to compete with MFCC-13 when using ResNet18 as well (VAR: 0.7323\hspace{1mm}\textless\hspace{1mm}MFCC-13: 0.7740, however not significantly, p=0.11). Finally, VAR performed better with the shallow model than with ResNet18 (shallow: 0.7583\hspace{1mm}\textgreater\hspace{1mm}ResNet18: 0.7323, however not significantly, p=0.54). The observation that the best scores of Tables \ref{tab:exp2} and \ref{tab:exp3} are close to each other shows that the following two options are equivalent: (i) taking the entire full-cycle and obtaining the representation with $N_t=256$, and (ii) dividing it into the six sub-phases, obtaining the representations with $N_t=64$ (the greater of the two options around 256$\div$6), and concatenating them side by side to form the overall representation.

%%%%%%%%%%%%%%%%%%%% 3333333333333333333333333333 %%%%%%%%%%%%%%%%%%%%%%%%%%%%%%
\begin{table}[b] 
\caption{F1-scores of the third experiment (testing the sub-phase temporal resolutions) with the shallow CNN.} \label{tab:exp3}
\centering
\begin{tabular}{r|cccc|}
\cline{2-5}
\multicolumn{1}{l|}{}                      & \multicolumn{4}{c|}{\textbf{$N_t$ ($N_{FFT}$)}}                                                                                                                                                                                    \\ \cline{2-5} 
\multicolumn{1}{l|}{}                      & \multicolumn{1}{c|}{\textbf{16 (4096)}}             & \multicolumn{1}{c|}{\textbf{32 (2048)}}             & \multicolumn{1}{c|}{\textbf{64 (1024)}}                                             & \textbf{128 (1024)}            \\ \hline
\multicolumn{1}{|r|}{\textbf{mfcc-13}}     & \multicolumn{1}{c|}{\cellcolor[HTML]{F3F3F3}0.7411} & \multicolumn{1}{c|}{\cellcolor[HTML]{F9F9F9}0.7764} & \multicolumn{1}{c|}{\cellcolor[HTML]{FFFFFF}{\color[HTML]{C00000} \textbf{0.8108}}} & \cellcolor[HTML]{E9E9E9}0.6823 \\ \hline
\multicolumn{1}{|r|}{\textbf{mfcc-26}}     & \multicolumn{1}{c|}{\cellcolor[HTML]{D4D4D4}0.5558} & \multicolumn{1}{c|}{\cellcolor[HTML]{9D9D9D}0.2346} & \multicolumn{1}{c|}{\cellcolor[HTML]{808080}0.0573}                                 & \cellcolor[HTML]{828282}0.0728 \\ \hline
\multicolumn{1}{|r|}{\textbf{mfcc-39}}     & \multicolumn{1}{c|}{\cellcolor[HTML]{9D9D9D}0.2317} & \multicolumn{1}{c|}{\cellcolor[HTML]{868686}0.0959} & \multicolumn{1}{c|}{\cellcolor[HTML]{8A8A8A}0.1185}                                 & \cellcolor[HTML]{898989}0.1132 \\ \hline
\multicolumn{1}{|r|}{\textbf{logmelsp-13}} & \multicolumn{1}{c|}{\cellcolor[HTML]{ECECEC}0.6999} & \multicolumn{1}{c|}{\cellcolor[HTML]{E2E2E2}0.6432} & \multicolumn{1}{c|}{\cellcolor[HTML]{DEDEDE}0.6159}                                 & \cellcolor[HTML]{DBDBDB}0.5987 \\ \hline
\multicolumn{1}{|r|}{\textbf{logmelsp-26}} & \multicolumn{1}{c|}{\cellcolor[HTML]{E7E7E7}0.6733} & \multicolumn{1}{c|}{\cellcolor[HTML]{DFDFDF}0.6217} & \multicolumn{1}{c|}{\cellcolor[HTML]{CACACA}0.5023}                                 & \cellcolor[HTML]{AEAEAE}0.3323 \\ \hline
\multicolumn{1}{|r|}{\textbf{logmelsp-39}} & \multicolumn{1}{c|}{\cellcolor[HTML]{E3E3E3}0.6463} & \multicolumn{1}{c|}{\cellcolor[HTML]{DFDFDF}0.6261} & \multicolumn{1}{c|}{\cellcolor[HTML]{B4B4B4}0.3713}                                 & \cellcolor[HTML]{A9A9A9}0.3015 \\ \hline
\multicolumn{1}{|r|}{\textbf{VAR}}         & \multicolumn{4}{c|}{\cellcolor[HTML]{F6F6F6}0.7583} \\ \hline
\end{tabular}
\end{table}
 
The results of the \textbf{fourth} experiment (testing the strategies for combining the sub-phases) are given in Table \ref{tab:exp4}. The score obtained with concatenation (0.8774) is not only above the scores obtained with GRU and GRU+Attention, but it is also the best score achieved among all the experiments conducted. Consequently, the best classifier design scenario turned out to be: \textit{using separate representations (obtained with MFCC-13, adaptive-length windowing, $N_t=64$) in multi-modal learning that performs direct concatenation of features for fusing the sub-phases}. The score achieved using this scenario (0.8774) is higher than those obtained using (i) full-cycle representation in single-modal learning (0.8104, Table \ref{tab:exp2}), (ii) joined representation in single-modal learning (0.8108, Table \ref{tab:exp3}), (iii) separate representation (MFCC-13, adaptive-length windowing, $N_t=64$) in multi-modal learning with GRU (0.8367, Table \ref{tab:exp4}), and (iv) separate representation (MFCC-13, adaptive-length windowing, $N_t=64$) in multi-modal learning with GRU+Attention (0.8408, Table \ref{tab:exp4}), however not significantly (p=0.18, p=0.17, p=0.33, and p=0.33, respectively). Briefly stated, (i) direct concatenation provided better performance than using more complex methods such as GRU or GRU+Atten-tion to combine the features, and (ii) concatenation in the feature space (i.e., learning the sub-phase characteristics individually, then combining what has been learned) improved the performance compared to concatenation in the representation space (i.e., learning the overall picture at once). The result that simple fusing outperformed the more complex methods can be attributed to the fact that learning sub-phase contributions was already handled at the level of feature extraction. Then, concatenating the features in the order of sub-phase timing was sufficient for diagnosis. 

%%%%%%%%%%%%%%%%%%%% 4444444444444444444444444 %%%%%%%%%%%%%%%%%%%%%%%%%%%%%%
\begin{table}[t] 
\caption{F1-scores of the fourth experiment (testing the strategies for combining
the sub-phases).} \label{tab:exp4}
\centering
\begin{tabular}{r|ccc|}
\cline{2-4}
\multicolumn{1}{c|}{}                      & \multicolumn{3}{c|}{\textbf{fusing type}}                                                                                                                                  \\ \cline{2-4} 
\multicolumn{1}{c|}{}                      & \multicolumn{1}{c|}{\textbf{concat}}                                                & \multicolumn{1}{c|}{\textbf{gru}}                   & \textbf{gru+attention}         \\ \hline
\multicolumn{1}{|r|}{\textbf{mfcc-13}}     & \multicolumn{1}{c|}{\cellcolor[HTML]{FFFFFF}{\color[HTML]{C00000} \textbf{0.8774}}} & \multicolumn{1}{c|}{\cellcolor[HTML]{EBEBEB}0.8367} & \cellcolor[HTML]{EDEDED}0.8408 \\ \hline
\multicolumn{1}{|r|}{\textbf{mfcc-26}}     & \multicolumn{1}{c|}{\cellcolor[HTML]{DADADA}0.8020}                                 & \multicolumn{1}{c|}{\cellcolor[HTML]{D4D4D4}0.7898} & \cellcolor[HTML]{F3F3F3}0.8530 \\ \hline
\multicolumn{1}{|r|}{\textbf{mfcc-39}}     & \multicolumn{1}{c|}{\cellcolor[HTML]{DCDCDC}0.8063}                                 & \multicolumn{1}{c|}{\cellcolor[HTML]{F6F6F6}0.8594} & \cellcolor[HTML]{EDEDED}0.8408 \\ \hline
\multicolumn{1}{|r|}{\textbf{logmelsp-13}} & \multicolumn{1}{c|}{\cellcolor[HTML]{B8B8B8}0.7341}                                 & \multicolumn{1}{c|}{\cellcolor[HTML]{A8A8A8}0.7015} & \cellcolor[HTML]{BFBFBF}0.7477 \\ \hline
\multicolumn{1}{|r|}{\textbf{logmelsp-26}} & \multicolumn{1}{c|}{\cellcolor[HTML]{8B8B8B}0.6422}                                 & \multicolumn{1}{c|}{\cellcolor[HTML]{D4D4D4}0.7898} & \cellcolor[HTML]{808080}0.6180 \\ \hline
\multicolumn{1}{|r|}{\textbf{logmelsp-39}} & \multicolumn{1}{c|}{\cellcolor[HTML]{878787}0.6341}                                 & \multicolumn{1}{c|}{\cellcolor[HTML]{939393}0.6583} & \cellcolor[HTML]{8F8F8F}0.6488 \\ \hline
\multicolumn{1}{|r|}{\textbf{VAR}}         & \multicolumn{1}{c|}{\cellcolor[HTML]{CACACA}0.7704}                                 & \multicolumn{1}{c|}{\cellcolor[HTML]{D5D5D5}0.7936} & \cellcolor[HTML]{D4D4D4}0.7906 \\ \hline
\end{tabular}
\end{table}

The results of the \textbf{fifth} experiment (testing the augmentation methods) are given in Table \ref{tab:exp5}. The reference values shown in the first column are from the previous (fourth) experiment (Table \ref{tab:exp4}). Among all tested scenarios, mixup augmentation yielded the best results when applied alone. However, augmentation in general was found to degrade the performance of the models overall. The best score achieved without augmentation (0.8774) is higher than its counterpart with mixup alone (0.8495), however not significantly (p=0.49). On the other hand, it is significantly better (p\hspace{1mm}\textless\hspace{1mm}0.05) than all scenarios with white noise infusion alone (WN[5]: 0.7429 with p=0.0195, WN[5,10]: 0.7675 with p=0.0350, and WN[5,10,15,20]: 0.7561 with p=0.0078). For comparison, mixup was also tested on the joined representation in single-modal learning (only for MFCC-13, adaptive-length windowing, $N_t=64$) and obtained a lower score (without mixup: 0.8108\hspace{1mm}\textgreater\hspace{1mm}with mixup: 0.7955), however not significantly (p=0.76). 

%%%%%%%%%%%%%%%%%%%% 555555555555555555555555555555 %%%%%%%%%%%%%%%%%%%%%%%%%%%%%%
\begin{table}[b]  
\caption{F1-scores of the fifth experiment (testing the augmentation methods). WN[x]: white noise infusion with SNR of x dB} \label{tab:exp5}
\centering
\begin{tabular}{crc|ccccccc|}
\cline{4-10}
\multicolumn{1}{l}{}                                         &                                           & \multicolumn{1}{l|}{}                                          & \multicolumn{7}{c|}{\textbf{augmentation scenario}}                                                                                                                                                                                                                                                                                                                                                                                                                                                                                                                      \\ \cline{3-10} 
\multicolumn{1}{l}{}                                         & \multicolumn{1}{r|}{}                     & \begin{turn}{90} \textbf{REF} \end{turn}                                                  & \multicolumn{1}{c|}{\begin{turn}{90} \textbf{M} \end{turn}}                 & \multicolumn{1}{c|}{\begin{turn}{90} \textbf{WN[5]} \end{turn}}       & \multicolumn{1}{c|}{\begin{turn}{90} \textbf{WN[5]+M} \end{turn}} & \multicolumn{1}{c|}{\begin{turn}{90} \textbf{WN[5,10]} \end{turn}}  & \multicolumn{1}{c|}{\begin{turn}{90} \textbf{WN[5,10]+M} \end{turn}} & \multicolumn{1}{c|}{\begin{turn}{90} \textbf{WN[5,10,15,20]} \end{turn}} & \begin{turn}{90} \textbf{WN[5,10,15,20]+M} \end{turn} \\ \hline
\multicolumn{1}{|c|}{}                                       & \multicolumn{1}{r|}{\textbf{mfcc13}}     & \cellcolor[HTML]{FFFFFF}{\color[HTML]{C00000} \textbf{0,8774}} & \multicolumn{1}{c|}{\cellcolor[HTML]{F5F5F5}0,8495} & \multicolumn{1}{c|}{\cellcolor[HTML]{D1D1D1}0,7429} & \multicolumn{1}{c|}{\cellcolor[HTML]{C9C9C9}0,7194}                                                     & \multicolumn{1}{c|}{\cellcolor[HTML]{D9D9D9}0,7675} & \multicolumn{1}{c|}{\cellcolor[HTML]{D2D2D2}0,7459}                                                         & \multicolumn{1}{c|}{\cellcolor[HTML]{D5D5D5}0,7561}         & \cellcolor[HTML]{DDDDDD}0,7783                                                                 \\ \cline{2-10} 
\multicolumn{1}{|c|}{}                                       & \multicolumn{1}{r|}{\textbf{lms13}} & \cellcolor[HTML]{CECECE}0,7341                                 & \multicolumn{1}{c|}{\cellcolor[HTML]{C9C9C9}0,7207} & \multicolumn{1}{c|}{\cellcolor[HTML]{909090}0,5502} & \multicolumn{1}{c|}{\cellcolor[HTML]{929292}0,5587}                                                     & \multicolumn{1}{c|}{\cellcolor[HTML]{959595}0,5651} & \multicolumn{1}{c|}{\cellcolor[HTML]{969696}0,5684}                                                         & \multicolumn{1}{c|}{\cellcolor[HTML]{808080}0,5030}         & \cellcolor[HTML]{969696}0,5689                                                                 \\ \cline{2-10} 
\multicolumn{1}{|c|}{\multirow{-3}{*}{\textbf{CONCAT}}}      & \multicolumn{1}{r|}{\textbf{VAR}}         & \cellcolor[HTML]{DADADA}0,7704                                 & \multicolumn{1}{c|}{\cellcolor[HTML]{DADADA}0,7703} & \multicolumn{1}{c|}{\cellcolor[HTML]{000000}-}                               & \multicolumn{1}{c|}{\cellcolor[HTML]{000000}-}                                                                                  & \multicolumn{1}{c|}{\cellcolor[HTML]{000000}-}                              & \multicolumn{1}{c|}{\cellcolor[HTML]{000000}-}                                                                                      & \multicolumn{1}{c|}{\cellcolor[HTML]{000000}-}                                       & \multicolumn{1}{c|}{\cellcolor[HTML]{000000}-}                                                                                               \\ \hline
\multicolumn{1}{|c|}{}                                       & \multicolumn{1}{r|}{\textbf{mfcc13}}     & \cellcolor[HTML]{F1F1F1}0,8367                                 & \multicolumn{1}{c|}{\cellcolor[HTML]{ECECEC}0,8243} & \multicolumn{1}{c|}{\cellcolor[HTML]{DFDFDF}0,7844} & \multicolumn{1}{c|}{\cellcolor[HTML]{C9C9C9}0,7202}                                                     & \multicolumn{1}{c|}{\cellcolor[HTML]{DEDEDE}0,7807} & \multicolumn{1}{c|}{\cellcolor[HTML]{E2E2E2}0,7924}                                                         & \multicolumn{1}{c|}{\cellcolor[HTML]{DCDCDC}0,7752}         & \cellcolor[HTML]{E0E0E0}0,7880                                                                 \\ \cline{2-10} 
\multicolumn{1}{|c|}{}                                       & \multicolumn{1}{r|}{\textbf{lms13}} & \cellcolor[HTML]{C3C3C3}0,7015                                 & \multicolumn{1}{c|}{\cellcolor[HTML]{BFBFBF}0,6897} & \multicolumn{1}{c|}{\cellcolor[HTML]{939393}0,5597} & \multicolumn{1}{c|}{\cellcolor[HTML]{828282}0,5096}                                                     & \multicolumn{1}{c|}{\cellcolor[HTML]{818181}0,5073} & \multicolumn{1}{c|}{\cellcolor[HTML]{969696}0,5690}                                                         & \multicolumn{1}{c|}{\cellcolor[HTML]{939393}0,5607}         & \cellcolor[HTML]{8F8F8F}0,5478                                                                 \\ \cline{2-10} 
\multicolumn{1}{|c|}{\multirow{-3}{*}{\textbf{GRU}}}         & \multicolumn{1}{r|}{\textbf{VAR}}         & \cellcolor[HTML]{E2E2E2}0,7936                                 & \multicolumn{1}{c|}{\cellcolor[HTML]{DDDDDD}0,7778} & \multicolumn{1}{c|}{\cellcolor[HTML]{000000}-}                              & \multicolumn{1}{c|}{\cellcolor[HTML]{000000}-}                                                                                  & \multicolumn{1}{c|}{\cellcolor[HTML]{000000}-}                            & \multicolumn{1}{c|}{\cellcolor[HTML]{000000}-}                                                                                       & \multicolumn{1}{c|}{\cellcolor[HTML]{000000}-}                                     & \multicolumn{1}{c|}{\cellcolor[HTML]{000000}-}                                                                                              \\ \hline
\multicolumn{1}{|c|}{}                                       & \multicolumn{1}{r|}{\textbf{mfcc13}}     & \cellcolor[HTML]{F2F2F2}0,8408                                 & \multicolumn{1}{c|}{\cellcolor[HTML]{F8F8F8}0,8593} & \multicolumn{1}{c|}{\cellcolor[HTML]{E7E7E7}0,8094} & \multicolumn{1}{c|}{\cellcolor[HTML]{DFDFDF}0,7851}                                                     & \multicolumn{1}{c|}{\cellcolor[HTML]{D0D0D0}0,7417} & \multicolumn{1}{c|}{\cellcolor[HTML]{DEDEDE}0,7828}                                                         & \multicolumn{1}{c|}{\cellcolor[HTML]{D4D4D4}0,7527}         & \cellcolor[HTML]{DBDBDB}0,7735                                                                 \\ \cline{2-10} 
\multicolumn{1}{|c|}{}                                       & \multicolumn{1}{r|}{\textbf{lms13}} & \cellcolor[HTML]{D3D3D3}0,7477                                 & \multicolumn{1}{c|}{\cellcolor[HTML]{C3C3C3}0,7024} & \multicolumn{1}{c|}{\cellcolor[HTML]{858585}0,5187} & \multicolumn{1}{c|}{\cellcolor[HTML]{828282}0,5095}                                                     & \multicolumn{1}{c|}{\cellcolor[HTML]{8A8A8A}0,5335} & \multicolumn{1}{c|}{\cellcolor[HTML]{A5A5A5}0,6128}                                                         & \multicolumn{1}{c|}{\cellcolor[HTML]{909090}0,5521}         & \cellcolor[HTML]{8F8F8F}0,5484                                                                 \\ \cline{2-10} 
\multicolumn{1}{|c|}{\multirow{-3}{*}{\textbf{GRU+A}}} & \multicolumn{1}{r|}{\textbf{VAR}}         & \cellcolor[HTML]{E1E1E1}0,7906                                 & \multicolumn{1}{c|}{\cellcolor[HTML]{DADADA}0,7690} & \multicolumn{1}{c|}{\cellcolor[HTML]{000000}-}                               & \multicolumn{1}{c|}{\cellcolor[HTML]{000000}-}                                                                                 & \multicolumn{1}{c|}{\cellcolor[HTML]{000000}-}                         & \multicolumn{1}{c|}{\cellcolor[HTML]{000000}-}                                                                                     & \multicolumn{1}{c|}{\cellcolor[HTML]{000000}-}                                      & \multicolumn{1}{c|}{\cellcolor[HTML]{000000}-}                                                                                               \\ \hline
\end{tabular}
\end{table}  

%%%%%%%%%%%%%%%%%%%%%%%% 66666666666666666666 %%%%%%%%%%%%%%%%%%%
\begin{table}[t]
\caption{Subject-based and cycle-based F1-scores, along with subject based accuracy, specificity and sensitivity rates. FCR: full-cycle representation, JR: joined representation, SR: separate representation, SR, MM: Multi-modal, SM: Single-modal} \label{tab:exp6}
\centering
\begin{tabular}{|c|c|c|c|c|c|c|c|c|}
\hline
\textbf{augmentation}  & \textbf{rep}     & \textbf{temp res} & \textbf{model}       & \textbf{Acc}                                                  & \textbf{Spec}                                                 & \textbf{Sens}                                                 & \textbf{F1(sbj)}                                                   & \textbf{F1(cyc)}                                             \\ \hline
-                      & mfcc-13          & t/p-256           & FCR, SM (Shal)                 & \cellcolor[HTML]{DADADA}0,820                                 & \cellcolor[HTML]{A7A7A7}0,878                                 & \cellcolor[HTML]{E9E9E9}0,733                                 & \cellcolor[HTML]{EBEBEB}0,765                                 & \cellcolor[HTML]{999999}0,715                                 \\ \hline
-             & \textbf{mfcc-13} & \textbf{adp-256}  & \textbf{FCR, SM (Shal)}        & \cellcolor[HTML]{FFFFFF}{\color[HTML]{C00000} \textbf{0,887}} & \cellcolor[HTML]{C7C7C7}0,922                        & \cellcolor[HTML]{FBFBFB}\textbf{0,833}                        & \cellcolor[HTML]{FFFFFF}{\color[HTML]{C00000} \textbf{0,855}} & \cellcolor[HTML]{D5D5D5}0,810                                 \\ \hline
-                      & mfcc-26          & t/p-256           & FCR, SM (RN18)                 & \cellcolor[HTML]{CCCCCC}0,793                                 & \cellcolor[HTML]{AFAFAF}0,889                                 & \cellcolor[HTML]{D9D9D9}0,650                                 & \cellcolor[HTML]{E0E0E0}0,716                                 & \cellcolor[HTML]{808080}0,674                                 \\ \hline
-                      & mfcc-26          & adp-256           & FCR, SM (RN18)                 & \cellcolor[HTML]{E1E1E1}0,833                                 & \cellcolor[HTML]{A7A7A7}0,878                                 & \cellcolor[HTML]{EFEFEF}0,767                                 & \cellcolor[HTML]{EFEFEF}0,786                                 & \cellcolor[HTML]{AAAAAA}0,741                                 \\ \hline
-                      & mfcc-13          & adp-64            & JR, SM   (Shal)      & \cellcolor[HTML]{F7F7F7}\textbf{0,873}                                 & \cellcolor[HTML]{AFAFAF}0,889                                 & \cellcolor[HTML]{FFFFFF}{\color[HTML]{C00000} \textbf{0,850}} & \cellcolor[HTML]{FCFCFC}\textbf{0,843}                                 & \cellcolor[HTML]{D5D5D5}0,811                                 \\ \hline
-                      & mfcc-13          & adp-64            & JR, SM   (RN18)      & \cellcolor[HTML]{DEDEDE}0,827                                 & \cellcolor[HTML]{808080}0,822                                 & \cellcolor[HTML]{FBFBFB}\textbf{0,833}                                 & \cellcolor[HTML]{F1F1F1}0,794                                 & \cellcolor[HTML]{BEBEBE}0,774                                 \\ \hline
-                      & VAR              & -            & JR, SM   (Shal)      & \cellcolor[HTML]{F0F0F0}0,860                                 & \cellcolor[HTML]{BFBFBF}0,911                                 & \cellcolor[HTML]{F2F2F2}0,783                                 & \cellcolor[HTML]{F6F6F6}0,817                                 & \cellcolor[HTML]{B4B4B4}0,758                                 \\ \hline
-                      & VAR              & -            & JR, SM   (RN18)      & \cellcolor[HTML]{DEDEDE}0,827                                 & \cellcolor[HTML]{B7B7B7}0,900                                 & \cellcolor[HTML]{E6E6E6}0,717                                 & \cellcolor[HTML]{EBEBEB}0,768                                 & \cellcolor[HTML]{A4A4A4}0,732                                 \\ \hline
-             & \textbf{mfcc-13} & \textbf{adp-64}   & \textbf{SR, MM   (con)} & \cellcolor[HTML]{9D9D9D}0,707                        & \cellcolor[HTML]{FFFFFF}{\color[HTML]{C00000} \textbf{1,000}} & \cellcolor[HTML]{929292}0,267                        & \cellcolor[HTML]{9F9F9F}0,421                        & \cellcolor[HTML]{FFFFFF}{\color[HTML]{C00000} \textbf{0,877}} \\ \hline
-                      & mfcc-13          & adp-64            & SR, MM   (gru)          & \cellcolor[HTML]{8E8E8E}0,680                                 & \cellcolor[HTML]{F7F7F7}\textbf{0,989}                                 & \cellcolor[HTML]{898989}0,217                                 & \cellcolor[HTML]{909090}0,351                                 & \cellcolor[HTML]{E5E5E5}0,837                                 \\ \hline
-                      & mfcc-13          & adp-64            & SR, MM   (gru+atten)        & \cellcolor[HTML]{959595}0,693                                 & \cellcolor[HTML]{F7F7F7}\textbf{0,989}                                 & \cellcolor[HTML]{8F8F8F}0,250                                 & \cellcolor[HTML]{999999}0,395                                 & \cellcolor[HTML]{E8E8E8}0,841                                 \\ \hline
M                      & mfcc-13          & adp-64            & JR, SM   (Shal)      & \cellcolor[HTML]{F0F0F0}0,860                                 & \cellcolor[HTML]{A7A7A7}0,878                                 & \cellcolor[HTML]{FBFBFB}\textbf{0,833}                                 & \cellcolor[HTML]{F8F8F8}0,826                                 & \cellcolor[HTML]{CBCBCB}0,795                                 \\ \hline
M                      & mfcc-13          & adp-64            & SR, MM   (con)          & \cellcolor[HTML]{A4A4A4}0,720                                 & \cellcolor[HTML]{F7F7F7}\textbf{0,989}                                 & \cellcolor[HTML]{9B9B9B}0,317                                 & \cellcolor[HTML]{ABABAB}0,475                                 & \cellcolor[HTML]{EDEDED}\textbf{0,850}                                 \\ \hline
WN[5]                 & mfcc-13          & adp-64            & SR, MM   (con)          & \cellcolor[HTML]{878787}0,667                                 & \cellcolor[HTML]{C7C7C7}0,922                                 & \cellcolor[HTML]{959595}0,283                                 & \cellcolor[HTML]{9B9B9B}0,405                                 & \cellcolor[HTML]{ABABAB}0,743                                 \\ \hline
M+WN[5]             & mfcc-13          & adp-64            & SR, MM   (con)          & \cellcolor[HTML]{A7A7A7}0,727                                 & \cellcolor[HTML]{CFCFCF}0,933                                 & \cellcolor[HTML]{AEAEAE}0,417                                 & \cellcolor[HTML]{BBBBBB}0,549                                 & \cellcolor[HTML]{9C9C9C}0,719                                 \\ \hline
WN[5,10]             & mfcc-13          & adp-64            & SR, MM   (con)          & \cellcolor[HTML]{808080}0,653                                 & \cellcolor[HTML]{EFEFEF}0,978                                 & \cellcolor[HTML]{808080}0,167                                 & \cellcolor[HTML]{808080}0,278                                 & \cellcolor[HTML]{BABABA}0,767                                 \\ \hline
M+WN[5,10]         & mfcc-13          & adp-64            & SR, MM   (con)          & \cellcolor[HTML]{8A8A8A}0,673                                 & \cellcolor[HTML]{BFBFBF}0,911                                 & \cellcolor[HTML]{9B9B9B}0,317                                 & \cellcolor[HTML]{A3A3A3}0,437                                 & \cellcolor[HTML]{ADADAD}0,746                                 \\ \hline
WN[5,10,15,20]     & mfcc-13          & adp-64            & SR, MM   (con)          & \cellcolor[HTML]{929292}0,687                                 & \cellcolor[HTML]{D7D7D7}0,944                                 & \cellcolor[HTML]{989898}0,300                                 & \cellcolor[HTML]{A2A2A2}0,434                                 & \cellcolor[HTML]{B3B3B3}0,756                                 \\ \hline
M+WN[5,10,15,20] & mfcc-13          & adp-64            & SR, MM   (con)          & \cellcolor[HTML]{929292}0,687                                 & \cellcolor[HTML]{B7B7B7}0,900                                 & \cellcolor[HTML]{A5A5A5}0,367                                 & \cellcolor[HTML]{ADADAD}0,484                                 & \cellcolor[HTML]{C1C1C1}0,778                                 \\ \hline
\end{tabular}
\end{table}

Finally, \textbf{subject-based} diagnostic capabilities of the classifiers were assessed, for the selected (prominent) options discussed throughout the experiments. For this purpose, the subject-based decisions were obtained by majority voting over the full-cycle decisions, and also by the alternative strategy based on the Bayesian decision rule (please refer to the penultimate paragraph of Section \ref{sec:ExperimentalDesign}). The results showed that simple majority voting outperformed the more complex approach offered by the Bayesian perspective, hence only the former will be reported here for brevity. Besides the F1-score, specificity, sensitivity and accuracy were also calculated to evaluate the subject-based performances, and all are reported in Table \ref{tab:exp6}. The cycle-based F1-scores addressed so far are given in the last column (after rounding to three decimal places to conserve space) for comparison. The table shows that: (i) the multi-modal scenarios increased the cycle-based F1-scores, and (ii) better diagnostic performance can be achieved when cycle decisions were combined to obtain subject diagnoses, although this has yet been observed only when single-modal scenarios were applied. According to the subject-based performance scores, the best F1-score (0.8547\hspace{1mm}$\approx$\hspace{1mm}0.855) and accuracy (0.8867\hspace{1mm} $\approx$\hspace{1mm}0.887) were obtained with the full-cycle representation (MFCC-13, adaptive-length windowing, $N_t=256$), the best sensitivity (0.850) was obtained with the joined representation (MFCC-13, adaptive-length windowing, $N_t=64$), and the best specificity (1.000) was obtained with the separate representation (MFCC-13, adaptive-length windowing, $N_t=64$) with direct feature concatenation for sub-phase fusion. The scenario with the highest subject-based specificity (1.000) also has the highest cycle-based F1-score (0.8774\hspace{1mm}$\approx$\hspace{1mm}0.877); however, it has a very low subject-based F1-score (0.421) and sensitivity (0.267), thus having a moderate accuracy (0.707). A notably low subject-based F1-score despite a maximal cycle-based F1-score is attributed to the pronounced cycle multiplicity observed in asthma patients compared to those with COPD. A large number of correctly predicted asthma cycles increased the cycle-based score significantly, while affecting the subject-based success negligibly in a positive way. On the other hand, correctly predicted COPD cycles that failed to reach the majority affected the cycle-based success negligibly in a negative way, but at the same time decreased the subject-based success significantly. Although subject-level scores were not so high when the multi-modal sub-phase fusion strategies were used, the improvement in cycle-level scores is promising. Theoretically, the more accurate the cycle-level predictions are, the more accurate the subject-level diagnosis would be. Searching for the optimum method (to be used instead of the two tested here) to combine the cycle decisions to arrive at subject decisions was kept outside the scope of this study.

\textit{Final Reflections:} While the vectorized VAR matrices were quite successful with classical machine learning methods as reported in \cite{Sen2021} (98\% subject-based accuracy), they were not so here in this study (around 85\% subject-based accuracy) with the CNN classifiers accepting them as inputs in their original matrix form. When the VAR matrices were vectorized and classified by the classical methods as in \cite{Sen2021}, each element of the matrix denotes an offset applied to the vector along the corresponding axis in the space, thereby having an individual effect on the separability of the two classes, and the classifier tries to learn the groups formed by the vectors. On the other hand, when they were used in their original matrix form and classified by convolutional models as in this study, neighbouring elements work together, and the classifier tries to learn the overall pattern of the matrix. Since an element in the VAR matrix is the model coefficient that represents the temporal relationship between two microphones, it can be interpreted to be more independent from its neighbors as compared to an element in the MFCC matrix that shows the variation of frequency characteristics over time. Having this nature, the VAR matrix apparently worked better (yielding the overall best performance) with the simpler classical classifier, whereas, with the same amount of data, the complex modern classifier made better use of the dependency inherent to the spectro-temporal representations. Until the data size can be increased enough to catch the accuracy rate of 98\% reported in \cite{Sen2021} by using the complex method combinations experimented in this study, it would be a sufficiently successful alternative to continue using classical methods with simpler input representations such as the VAR model matrix.

Overall, MFCC was more successful than both log-spectrogram and log-mel spectrogram. Post-processing the log-mel spectrogram by DCT to obtain the MFCC seems to enhance the discriminative power in distinguishing asthma and COPD, implying that the harmonic pattern of frequency components is the distinguishing factor, rather than the exact spectral content. Physiological justification can only be provided by acoustic testing using a realistic lung model that mimics the airway tissues that are altered in different ways by asthma and COPD. MFCC is already known to perform successfully in speech processing, for which it was first introduced, and since both speech and pulmonary sounds are processed by the same auditory perception mechanism, its success here was not surprising, though its notable superiority over VAR was. The performance of the VAR model was generally between that of MFCC and log-mel spectrogram, however it was clearly below expectations, especially considering that the matrices (which are independent of timing or amplitude variations across the recordings and are always the same size) are numerical representations of spatio-temporal relationships between the channels and performed well in \cite{Sen2021}. This result showed at least that the pairwise correlations of channels (reflecting themselves in the VAR coefficients) are discriminative to the extent that classical machine learning methods can handle, however the relationships between the pairwise correlations (spatial alignment of the VAR coefficients) are not as discriminative as the spatial information provided by the MFCC matrix exploited best by CNN. 

Finally, the results show that augmentation methods did not help better learning. These methods, if they worked, could be consulted in order to improve training as it is difficult to build large amounts of annotated data in our case, because the measurements: (i) require attachement of multi sensor equipment, which can be time-consuming, (ii) are usually repeated to increase the chance of data usability, (iii) necessitate trained personnel due to sensitive requirements for proper attachement, and (iv) necessitate dedicated personnel and physical space, which is difficult to have in the busy hospital routine. The reason that augmentation did not work can be attributed to the fact that the two dimensional inputs are direct numerical representations of the acoustic information content affected by distinct clinical conditions, rather than, e.g., photographs that may deviate from a reference image due to rotation, scale, or illumination. In our case, improved learning can only be achieved as new annotated data are collected and used to update the models. 

\section{Conclusion} 
\label{sec:Conclusion}
In diagnostic classification of asthma and COPD using the CNN models tested, MFCC outperformed the other two spectro-temporal representations, namely, the log-mel spectrogram and log-spectrogram. Thirteen coefficients were the best to convey the discriminative information. The prominent aspect of MFCC is the DCT applied after mel filtering, implying that the discriminative information lies in the harmonic frequency patterns of asthma and COPD sounds, rather than in their exact frequency content. The VAR model did not perform as expected, yielding lower success than both MFCC and its own prior performance with classical machine learning classifiers. The augmentation methods tested here did not improve the performance, as would be expected given that numerical representations computed from acoustic clinical data are being classified.

To equalize the temporal sizes of the spectro-temporal representations, trimming/padding and adaptive-length windowing scenarios are equally applicable; nevertheless, between the two, the latter is recommended due to its higher average success. A 64-point time resolution per sub-phase or a 256-point time resolution per full-cycle is recommended for single-modal systems. The best performance was achieved by the multi-modal approach where sub-phase fusion is done by direct concatenation in the feature space. The result that simple fusing outperformed more complex methods like GRU and GRU+Attention can be attributed to the fact that learning sub-phase contributions was already handled at the level of feature extraction and concatenating them in the order of their timing was sufficient. Overall, the performance improvement provided by the multi-modal approach suggests that sub-phases should indeed be handled separately for improved diagnostic classification of pulmonary sounds.

Having in mind that it is necessary to find an improved (than majority voting) way of combining the full-cycle decisions to obtain subject-based diagnoses, the best methodology suggested by this work is using separate representations of sub-phases via six MFCC-13 matrices each having 64-point time resolution by adaptive-length windowing, processed by the multi-modal network with direct concatenation of features in the feature space (F1-score: 0.8774).

Until a reliable dataset including a large amount of recordings from a large number of different subjects with various conditions (singular or comorbid diagnoses) that can be used to train deep learning algorithms for reliable clinical performance, shallow networks can be said to be preferable over the deeper ones, and simpler sub-phase fusing can be said to be preferable over the more sophisticated methods. Furthermore, classical machine learning algorithm can still be preferred, if in accordance with the complexity of the inputs and the requirements of the application.

Improved learning can be achieved as new annotated data enters into the system and the deep learning classifiers are trained on the updated data set, and by optimizing the method to combine the full-cycle decisions to obtain subject-based decisions.

\section*{CRediT authorship contribution statement}
\noindent \textbf{Ipek Sen:} Conceptualization, Methodology, Validation, Investigation, Resources, Data Curation, Writing - Original Draft (80\%), Visualization, Supervision, Project administration (principal investigator), Funding acquisition. \textbf{Ozgur Ozdemir:} Methodology, Software, Validation, Formal analysis, Investigation, Resources, Writing - Original Draft (15\%), Visualization. \textbf{Elena Battini Sonmez:} Writing - Original Draft (5\%), Project administration (team member as researcher)

\section*{Ethics Statement}
The authors declare that all procedures performed in this study involving human participants had the approval (No: B.30.2.IST.0.02.00.01/YI-1157) of the Second Ethical Committee of Clinical Research, Istanbul Medical Faculty, Istanbul University, and were in accordance with the Declaration of Helsinki. An informed consent was taken from each subject before participating in the study.

\section*{Funding}
This work was funded by Istanbul Bilgi University BAP (no: 2020.02.005).

\section*{Declaration of Competing Interest}
The authors declare that they have no known competing financial interests or personal relationships that could have appeared to influence the work reported in this paper.

\section*{Data Availability}
The data used in this study is not publicly available and requires special permission.

\section*{Declaration of Generative AI and AI-assisted Technologies in the Writing Process}
During the preparation of this work, the authors used DeepL for translation and grammatical improvement. The tool was used sparingly, and the authors carefully reviewed and edited the content as needed. The authors take full responsibility for the content of this publication.

\bibliographystyle{unsrt}  
\bibliography{main}

@article{sengupta2016,
title = {Lung sound classification using cepstral-based statistical features},
journal = {Computers in Biology and Medicine},
volume = {75},
pages = {118-129},
year = {2016},
author = {Nandini Sengupta and Md Sahidullah and Goutam Saha},
}

@article{wang2025,
  author={Wang, Fan and Gao, Jiacheng and Wang, Ying and Huang, Guoheng and Yuan, Xiaochen},
  journal={IEEE Access}, 
  title={Hybrid Dual-Input Model for Respiratory Sound Classification With Mel Spectrogram and Waveform}, 
  year={2025},
  volume={13},
  number={},
  pages={80971-80980},
}

@article{zulfiqar2021,
  author    = {Zulfiqar, F. and others},
  title     = {Abnormal Respiratory Sounds Classification Using Deep CNN},
  journal   = {Frontiers in Medicine (Lausanne)},
  volume    = {8},
  pages     = {714811},
  year      = {2021},
}

@article{asatani2021,
title = {Classification of respiratory sounds using improved convolutional recurrent neural network},
journal = {Computers \& Electrical Engineering},
volume = {94},
pages = {107367},
year = {2021},
author = {Naoki Asatani and Tohru Kamiya and Shingo Mabu and Shoji Kido},
}

@article{petmezas2022,
AUTHOR = {Petmezas, Georgios and Cheimariotis, Grigorios-Aris and Stefanopoulos, Leandros and Rocha, Bruno and Paiva, Rui Pedro and Katsaggelos, Aggelos K. and Maglaveras, Nicos},
TITLE = {Automated Lung Sound Classification Using a Hybrid CNN-LSTM Network and Focal Loss Function},
JOURNAL = {Sensors},
VOLUME = {22},
YEAR = {2022},
NUMBER = {3},
ARTICLE-NUMBER = {1232},
}

@article{kim2021,
  author    = {Yoonjoo Kim and YunKyong Hyon and Sung Soo Jung and Sunju Lee and Geon Yoo and Chaeuk Chung and Taeyoung Ha},
  title     = {Respiratory sound classification for crackles, wheezes, and rhonchi in the clinical field using deep learning},
  journal   = {Scientific Reports},
  volume    = {11},
  number    = {1},
  pages     = {17186},
  year      = {2021},
}

@article{zhang2024,
title = {Research on lung sound classification model based on dual-channel CNN-LSTM algorithm},
journal = {Biomedical Signal Processing and Control},
volume = {94},
pages = {106257},
year = {2024},
author = {Yipeng Zhang and Qiong Huang and Wenhui Sun and Fenlan Chen and Dongmei Lin and Fuming Chen},
}

@article{shehab2024,
  author    = {Sara A. Shehab and Kamel K. Mohammed and Ashraf Darwish and Aboul Ella Hassanien},
  title     = {Deep learning and feature fusion-based lung sound recognition model to diagnoses the respiratory diseases},
  journal   = {Soft Computing},
  volume    = {28},
  pages     = {11667--11683},
  year      = {2024},
}

@article{wang2024,
  author    = {Zhaoping Wang and Zhiqiang Sun},
  title     = {erformance evaluation of lung sounds classification using deep learning under variable parameters},
  journal   = {EURASIP Journal on Advances in Signal Processing},
  volume    = {2024},
  pages     = {51},
  year      = {2024},
}

@article{jung2021,
  title={Efficiently Classifying Lung Sounds through Depthwise Separable CNN Models with Fused STFT and MFCC Features},
  author={Jung, Shing-Yun and Liao, Chia-Hung and Wu, Yu-Sheng and Yuan, Shyan-Ming and Sun, Chuen-Tsai},
  journal={Diagnostics (Basel)},
  volume={11},
  number={4},
  pages={732},
  year={2021},
}

@article{kim2025,
  title={Enhanced Respiratory Sound Classification Using Deep Learning and Multi-Channel Auscultation},
  author={Kim, Yeonkyeong and Kim, Kyu Bom and Leem, Ah Young and Kim, Kyuseok and Lee, Su Hwan},
  journal={Journal of Clinical Medicine},
  volume={14},
  number={15},
  pages={5437},
  year={2025},
}

@article{haider2019,
  title={Respiratory Sound Based Classification of Chronic Obstructive Pulmonary Disease: a Risk Stratification Approach in Machine Learning Paradigm},
  author={Haider, Nishi Shahnaj and Singh, Bikesh Kumar and Periyasamy, R. and Behera, Ajoy K.},
  journal={Journal of Medical Systems},
  volume={43},
  number={255},
  year={2019},
}

@article{palaniappan2014,
  title={A comparative study of the SVM and K-NN machine learning algorithms for the diagnosis of respiratory pathologies using pulmonary acoustic signals},
  author={Palaniappan, Rajkumar and Sundaraj, Kenneth and Sundaraj, Sebastian},
  journal={BMC Bioinformatics},
  volume={15},
  number={223},
  year={2014},
}

@misc{fraiwan2021dataset,
  author       = {Fraiwan, Mohammad and Fraiwan, Luay and Khassawneh, Basheer and Ibnian, Ali},
  title        = {A dataset of lung sounds recorded from the chest wall using an electronic stethoscope},
  year         = {2021},
  publisher    = {Mendeley Data},
  version      = {V3},
  url          = {https://data.mendeley.com/datasets/jwyy9np4gv/3}
}

@INPROCEEDINGS{koshta2023,
  author={Koshta, Vaibhav and Singh, Bikesh Kumar and Behera, Ajoy K. and Ganga, Ranganath T.},
  booktitle={2023 11th International Conference on Intelligent Systems and Embedded Design (ISED)}, 
  title={Classification of Asthma, COPD and Healthy Lung Sounds Using Fourier Bessel Series Expansion in Machine Learning and Deep Learning Paradigm}, 
  year={2023},
  volume={},
  number={},
  pages={1-6},
}

@article{haider2022,
  title={Computerized lung sound based classification of asthma and chronic obstructive pulmonary disease (COPD)},
  author={Haider, Nishi Shahnaj and Behera, AK},
  journal={Biocybernetics and Biomedical Engineering},
  volume={42},
  number={1},
  pages={42--59},
  year={2022},
  publisher={Elsevier}
}

@article{khodabakhshi2017,
title = {The attractor recurrent neural network based on fuzzy functions: An effective model for the classification of lung abnormalities},
journal = {Computers in Biology and Medicine},
volume = {84},
pages = {124-136},
year = {2017},
author = {Mohammad Bagher Khodabakhshi and Mohammad Hassan Moradi},
}

@article{hossain2019,
  title={A comprehensive survey of deep learning for image captioning},
  author={Hossain, MD Zakir and Sohel, Ferdous and Shiratuddin, Mohd Fairuz and Laga, Hamid},
  journal={ACM Computing Surveys (CsUR)},
  volume={51},
  number={6},
  pages={1-36},
  year={2019}
}

@article{jha2023,
  title={Extracting low-dimensional psychological representations from convolutional neural networks},
  author={Jha, Aditi and Peterson, Joshua C and Griffiths, Thomas L},
  journal={Cognitive Science},
  volume={47},
  number={1},
  pages={e13226},
  year={2023}
}

@article{farrajota2019,
  title={Human action recognition in videos with articulated pose information by deep networks},
  author={Farrajota, Miguel and Rodrigues, Jo{\~a}o MF and du Buf, JM Hans},
  journal={Pattern Analysis and Applications},
  volume={22},
  pages={1307--1318},
  year={2019}
}

@article{iqbal2021,
  title={Automated multi-class classification of skin lesions through deep convolutional neural network with dermoscopic images},
  author={Iqbal, Imran and Younus, Muhammad and Walayat, Khuram and Kakar, Mohib Ullah and Ma, Jinwen},
  journal={Computerized medical imaging and graphics},
  volume={88},
  pages={101843},
  year={2021}
}

@inproceedings{koike2021,
  title={Transferring cross-corpus knowledge: An investigation on data augmentation for heart sound classification},
  author={Koike, Tomoya and Qian, Kun and Schuller, Bj{\"o}rn W and Yamamoto, Yoshiharu},
  booktitle={2021 43rd Annual International Conference of the IEEE Engineering in Medicine \& Biology Society (EMBC)},
  pages={1976-1979},
  year={2021}
}

@article{bai2018,
  title={An empirical evaluation of generic convolutional and recurrent networks for sequence modeling},
  author={Bai, Shaojie and Kolter, J Zico and Koltun, Vladlen},
  journal={arXiv preprint arXiv:1803.01271},
  year={2018}
}

@article{zhang2017,
  title={mixup: Beyond empirical risk minimization},
  author={Zhang, Hongyi and Cisse, Moustapha and Dauphin, Yann N and Lopez-Paz, David},
  journal={arXiv preprint arXiv:1710.09412},
  year={2017}
}

@article{siriwardhana2020,
  title={Jointly fine-tuning" bert-like" self supervised models to improve multimodal speech emotion recognition},
  author={Siriwardhana, Shamane and Reis, Andrew and Weerasekera, Rivindu and Nanayakkara, Suranga},
  journal={arXiv preprint arXiv:2008.06682},
  year={2020}
}

@article{bahdanau2014,
  title={Neural machine translation by jointly learning to align and translate},
  author={Bahdanau, Dzmitry and Cho, Kyunghyun and Bengio, Yoshua},
  journal={arXiv preprint arXiv:1409.0473},
  year={2014}
}

@article{simonyan2014,
  title={Very deep convolutional networks for large-scale image recognition},
  author={Simonyan, Karen and Zisserman, Andrew},
  journal={arXiv preprint arXiv:1409.1556},
  year={2014}
}

@article{zagoruyko2016,
  title={Wide residual networks},
  author={Zagoruyko, Sergey and Komodakis, Nikos},
  journal={arXiv preprint arXiv:1605.07146},
  year={2016}
}

@inproceedings{huang2017,
  title={Densely connected convolutional networks},
  author={Huang, Gao and Liu, Zhuang and Van Der Maaten, Laurens and Weinberger, Kilian Q},
  booktitle={Proceedings of the IEEE conference on computer vision and pattern recognition},
  pages={4700-4708},
  year={2017}
}

@inproceedings{he2016,
  title={Deep residual learning for image recognition},
  author={He, Kaiming and Zhang, Xiangyu and Ren, Shaoqing and Sun, Jian},
  booktitle={Proceedings of the IEEE conference on computer vision and pattern recognition},
  pages={770-778},
  year={2016}
}

@article{Messner2020,
  title = {Multi-channel lung sound classification with convolutional recurrent neural networks},
  journal = {Computers in Biology and Medicine},
  volume = {122},
  pages = {103831},
  year = {2020},
  author = {Elmar Messner and Melanie Fediuk and Paul Swatek and Stefan Scheidl and Freyja-Maria Smolle-Jüttner and Horst Olschewski and Franz Pernkopf}
}

@article{Srivastava2021,
  title={Deep learning based respiratory sound analysis for detection of chronic obstructive pulmonary disease},
  author={Srivastava, A.and Jain, S. and Miranda, R. and Patil, S. and Pandya, S. and Kotecha, K.},
  journal={PeerJ. Computer science},
  year={2021},
  volume={7},
  number={e369}
}

@article{Tinkelman2006,
  title = {Symptom-based questionnaire for differentiating COPD and asthma},
  author = {Tinkelman, {David G.} and Price, {David B.} and Nordyke, {Robert J.} and Halbert, {R. J.} and Sharon Isonaka and Dmitry Nonikov and Juniper, {Elizabeth F.} and Daryl Freeman and Thomas Hausen and Levy, {Mark L.} and Anders Ostrem and {van der Molen}, Thys and {van Schayck}, {Constant P.}},
  year = {2006},
  volume = {73},
  pages = {296-305},
  journal = {Respiration},
  number = {3}
}

@article{VanSchayck1996,
  title={Diagnosis of asthma and chronic obstructive pulmonary disease in general practice.},
  author={Van Schayck, CP},
  journal={The British Journal of General Practice},
  volume={46},
  number={404},
  pages={193-197},
  year={1996},
  publisher={Royal College of General Practitioners}
}

@article{Bellia2003,
  title={Aging and disability affect misdiagnosis of COPD in elderly asthmatics: the SARA study},
  author={Bellia, Vincenzo and Battaglia, Salvatore and Catalano, Filippo and Scichilone, Nicola and Incalzi, Raffaele Antonelli and Imperiale, Claudio and Rengo, Franco},
  journal={Chest},
  volume={123},
  number={4},
  pages={1066--1072},
  year={2003}
}

@article{Burrows1991,
  title={Characteristics of asthma among elderly adults in a sample of the general population},
  author={Burrows, Benjamin and Barbee, Robert A and Cline, Martha G and Knudson, Ronald J and Lebowitz, Michael D},
  journal={Chest},
  volume={100},
  number={4},
  pages={935-942},
  year={1991}
}

@article{ATS1987,
  title={Standards for the diagnosis and care of patients with chronic obstructive pulmonary disease (COPD) and asthma. Official statement},
  author={American Thoracic Society},
  journal={Am Rev Respir Dis},
  volume={136},
  pages={225-224},
  year={1987}
}

@article{VanWheel2002,
  title={Underdiagnosis of asthma and COPD: is the general practitioner to blame?},
  author={van Wheel, C.},
  journal={Monaldi Archives for Chest Disease},
  volume={57},
  number={1},
  pages={65-68},
  year={2002}
}

@article{Malmberg1995,
  title={Significant differences in flow standardised breath sound spectra in patients with chronic obstructive pulmonary disease, stable asthma, and healthy lungs},
  author={Malmberg, LP and Pesu, L and Sovijarvi, AR},
  journal={Thorax},
  volume={50},
  number={12},
  pages={1285-1291},
  year={1995}
}

@article{Sen2021,
  title={Differential Diagnosis of Asthma and COPD Based on Multivariate Pulmonary Sounds Analysis},
  author={Sen, I and Saraclar, M and Kahya YP},
  journal={IEEE Trans Biomed Eng.},
  volume={68},
  number={5},
  pages={1601-1610},
  year={2021}
}

@INPROCEEDINGS{Islam2018,
  author={Islam, Md. Ariful and Bandyopadhyaya, Irin and Bhattacharyya, Parthasarathi and Saha, Goutam},
  booktitle={2018 International Conference on Communication and Signal Processing (ICCSP)}, 
  title={Classification of Normal, Asthma and COPD Subjects Using Multichannel Lung Sound Signals}, 
  year={2018},
  pages={290-294}
}

@book{Gavriely1995,
  title="Breath Sounds Methodology",
  author="Noam Gavriely",
  year=1995,
  publisher="CRC Press",
  address="Florida"
}

@article{Nath1974,
  title={Significant differences in flow standardised breath sound spectra in patients with chronic obstructive pulmonary disease, stable asthma, and healthy lungs},
  author={Nath, AR and Capel, LH},
  journal={Thorax},
  volume={29},
  number={2},
  pages={223-227},
  year={1974}
}

@article{Sovijarvi2000p591,
  title={Significant differences in flow standardised breath sound spectra in patients with chronic obstructive pulmonary disease, stable asthma, and healthy lungs},
  author={Sovijarvi, ARA and Malmberg, LP and Charbonneau, G. and  Vanderschoot, J. and Dalmasso, F. and Sacco, C. and Rossi, M. and Earis, JE},
  journal={European Respiratory Review},
  volume={10},
  number={77},
  pages={591-596},
  year={2000}
}

@book{Douros2018,
  title={Breath Sounds},
  author={Douros,K. and Grammeniatis, V.and Loukou, I.},
  year={2018},
  publisher={Priftis, K. and Hadjileontiadis, L.and Everard M. (eds) Springer Cham}
}

@article{Piirila1991,
  title={Crackles in patients with fibrosing alveolitis, bronchiectasis, COPD, and heart failure},
  author={Piirilä, P and Sovijärvi, AR and Kaisla, T and Rajala, HM and Katila T.},
  journal={Chest},
  volume={99},
  number={},
  pages={1076-1083},
  year={1991}
}

@INPROCEEDINGS{Sen2005,
  author={Sen, I. and Kahya, Y.P.},
  booktitle={2005 IEEE Engineering in Medicine and Biology 27th Annual Conference}, 
  title={A Multi-Channel Device for Respiratory Sound Data Acquisition and Transient Detection}, 
  year={2005},
  volume={},
  number={},
  pages={6658-6661}
  }

@article{pavlitskaya2022,
  title={Measuring Overfitting in Convolutional Neural Networks using Adversarial Perturbations and Label Noise},
  author={Pavlitskaya, Svetlana and Oswald, Jo{\"e}l and Z{\"o}llner, J Marius},
  journal={arXiv preprint arXiv:2209.13382},
  year={2022}
}

@article{hangaard2017causes,
  title={Causes of misdiagnosis of chronic obstructive pulmonary disease: a systematic scoping review},
  author={Hangaard, Stine and Helle, Tina and Nielsen, Carl and Hejlesen, Ole K},
  journal={Respiratory medicine},
  volume={129},
  pages={63-84},
  year={2017},
  publisher={Elsevier}
}

@article{pistelli2011practical,
  title={Practical management problems of stable chronic obstructive pulmonary disease in the elderly},
  author={Pistelli, Riccardo and Ferrara, Letizia and Misuraca, Clementina and Bustacchini, Silvia},
  journal={Current opinion in pulmonary medicine},
  volume={17},
  pages={S43--S48},
  year={2011}
}

@article{sovijarvi2000p597,
  title = {Definition of Terms for Applications of Respiratory Sounds},
  author = {Sovijärvi, Anssi and Dalmasso, F. and Vanderschoot, J. and Malmberg, Leo and Righini, G. and Stoneman, S.A.T.},
  journal = {Eur Respir Rev},
  volume = {10},
  number={77},
  pages = {597-610},
  year = {2000},
}

@article{ATS,
  title = "American {T}horacic {S}ociety, {U}pdated nomenclature for membership relation",
  author = "",
  journal = {ATS News},
  volume = {3},
  pages = {5-6},
  year = 1977,
}

@article{Sen2013,
  title = {Exploring an optimal vector autoregressive model for multi-channel pulmonary sound data},
  author = {Ipek Sen and Murat Saraclar and Yasemin P. Kahya},
  journal = {Computer Methods and Programs in Biomedicine},
  volume = {111},
  number = {3},
  pages = {550-560},
  year = {2013},
}

@article{Davis1980,
  author={Davis, S. and Mermelstein, P.},
  journal={IEEE Transactions on Acoustics, Speech, and Signal Processing}, 
  title={Comparison of parametric representations for monosyllabic word recognition in continuously spoken sentences}, 
  year={1980},
  volume={28},
  number={4},
  pages={357-366},
}

@inproceedings{Logan2000,
  title={Mel frequency cepstral coefficients for music modeling.},
  author={Logan, Beth and others},
  booktitle={Ismir},
  volume={270},
  pages={11},
  year={2000},
  organization={Plymouth, MA}
}
\end{document}